\documentclass[twocolumn,twocolappendix]{aastex631}
\usepackage[utf8]{inputenc}
\usepackage{float}
\usepackage{amssymb,natbib,graphicx,color,ctable, amsmath,array}
\usepackage{hyperref}
\usepackage{mathtools}
\usepackage{soul}
\hypersetup{
    colorlinks=true,
    linkcolor=blue,
    citecolor=blue,
    filecolor=blue,
    urlcolor=blue
}
\usepackage[flushleft]{threeparttable}

\shortauthors{Sternberg et al.}

\begin{document}

\defcitealias{Sternberg2014}{S14}
\defcitealias{Bialy2016}{BS16}

\title{HI in Molecular Clouds: Irradiation by FUV plus Cosmic Rays}


\author{Amiel Sternberg}
\affiliation{School of Physics and Astronomy, Tel Aviv University, Ramat Aviv 69978, Israel}
\affiliation{Center for Computational Astrophysics, Flatiron Institute, 162 5th Ave., New York, NY, 10010}
\affiliation{Max-Planck-Institut f\"ur extraterrestrische Physik (MPE), Giessenbachstr., 85748 Garching, FRG}

\author{Shmuel Bialy}
\affiliation{Physics Department, Technion, Haifa, 3200003, Israel}

\author{Alon Gurman}
\affiliation{School of Physics \& Astronomy, Tel Aviv University, Ramat Aviv 69978, Israel}

\date{Accepted -. Received: Aug 26, 2023}



\begin{abstract}
We extend the analytic theory presented by \cite{Sternberg2014} and \cite{Bialy2016} for the production of atomic hydrogen (HI) via FUV photodissociation at the boundaries of dense interstellar molecular (H$_2$) clouds, to also include the effects of penetrating (low-energy) cosmic-rays for the growth of the total HI column densities. We compute the steady-state abundances of the HI and H$_2$ in one-dimensional gas slabs in which the FUV photodissociation rates are reduced by depth-dependent H$_2$ self-shielding and dust absorption, and in which the cosmic-ray ionization rates are either constant or reduced by transport effects. The solutions for the HI and H$_2$ density profiles and the integrated HI columns, depend primarily on the ratios $I_{\rm UV}/Rn$ and $\zeta/Rn$, where $I_{\rm UV}$ is the intensity of the photodissociating FUV field, $\zeta$ is the H$_2$ cosmic-ray ionization rate, $n$ is the hydrogen gas density, and $R$ is the dust-surface H$_2$ formation rate coefficient. We present computations for a wide range of FUV field strengths, cosmic-ray ionization rates, and dust-to-gas ratios. We develop analytic expressions for the growth of the HI column densities. For Galactic giant molecular clouds (GMCs) with multiphased (warm/cold) HI envelopes, the interior cosmic-ray zones will dominate the production of the HI only if $\zeta \gtrsim 4.5\times 10^{-16} \times (M_{\rm GMC}/10^6 \ M_{\odot})^{-1/2}$~s$^{-1}$, where $M_{\rm GMC}$ is the GMC mass, and including attenuation of the cosmic-ray fluxes. For most Galactic GMCs and conditions, FUV photodissociation dominates over cosmic-ray ionization for the production of the HI column densities. Furthermore, the cosmic-rays do not affect the HI-to-H$_2$ transition points.
\end{abstract}


\vspace{0.5cm}




\section{Introduction}

The compression of diffuse and warm interstellar atomic hydrogen (HI) gas into dense cold giant molecular (H$_2$) clouds (GMCs) is associated with radiative cooling, gravitational collapse, chemical complexity, and galaxy- star- and planet-formation across cosmic time \citep[hereafter \citetalias{Sternberg2014}]{McKee2007,Tacconi2020,Chevance2022,Sternberg2014}.
Much of the cold ($\lesssim 500$~K) HI observed via 21~cm emissions and absorptions in the interstellar medium (ISM) of galaxies is produced in photodissociation regions (PDRs) in the atomic to molecular (HI-to-H$_2$) transition layers of the dense star-forming molecular clouds
\citep{Allen1986,Heiner2011,Walter2008,Bialy2017,Schruba2018,Saintonge2022}. 

In recent years, hydrodynamical simulations of ever increasing sophistication have been incorporating the coupled radiative transfer and chemical processes necessary for appropriate modeling of the cold HI and H$_2$ components of the ISM 
\citep{Bialy2017,Nickerson2018,Inoue2020,Seifried2022,Katz2022,Gebek2023,Hu2021,Hu2022,Hopkins2023,KimB2023,Gurman2023}. Semianalytic methodology, including ``classical" one-dimensional (1D) PDR modeling \citep{Tielens1985,vanDishoeck1988,Sternberg1989,Wolfire2022} remain essential tools for interpreting observations, and for understanding and post-processing the results of the hydrodynamical simulations \citep{Levrier2012,Bialy2016,Rollig2022,Pound2023,
Kim2023,Bisbas2023}.

H$_2$ photodissociation by far-ultraviolet (FUV, $\sim$ 1000 \AA) radiation is limited by dust absorption to typical hydrogen gas column densities of $\sim 10^{21}$~cm$^{-2}$ (or gas surface densities $\Sigma_{\rm gas} \sim 11$~M$_\odot$~pc$^{-2}$ including helium) with temperatures $\sim 100$~K. At greater cloud depths a residual (but still significant) abundance of ultra-cold ($\sim 20$~K) atomic hydrogen may be maintained by low-energy ($\lesssim$~1 Gev) cosmic ray proton bombardment \citep{Spitzer1968,Solomon1971,Dalgarno2006,Gabici2022}, and observable as ``HI narrow self-absorption" (HINSA) features in 21 cm line profiles \citep{Knapp1974,Li2003,Goldsmith2007,Seifried2022}.

What are the relative contributions of FUV photodissociation and cosmic-ray bombardment to the production of HI in typical molecular clouds in star-forming galaxies? In \citetalias{Sternberg2014} and 
\citet[hereafter \citetalias{Bialy2016}]{Bialy2016}
we presented numerical and analytic theory for the HI column densities produced by photodissociation in the HI-to-H$_2$ transition layers in optically thick and dusty PDRs, but with the exclusion of cosmic-rays. In \cite{Bialy2015} we investigated the HI/H$_2$ balance in low-metallicity cloud interiors dominated by cosmic-ray processes, but with no FUV. In \cite{Sternberg2021} we did consider combined FUV and cosmic-ray irradiation, but for dust-{\it free} systems in which cosmic-ray ionization, rather than dust catalysis, drives a gas-phase conversion of HI to H$_2$, and in which the attenuation of the photodissocation rate is via pure H$_2$ absorption line self-shielding. Such dust-free PDRs may be relevant for young Universe conditions at the epoch of first star-formation. In this paper, we extend the analytic theory we presented in \citetalias{Sternberg2014} and 
\citetalias{Bialy2016} for {\it dusty} clouds, to also include cosmic-ray removal of the H$_2$ and the associated production of residual HI in the extended molecular cloud interiors. This in addition to direct photodissociation in the cloud surface PDRs.

In \S 2.1 and 2.2 we write down our basic HI/H$_2$ formation-destruction equation that  includes a term for cosmic-ray removal of H$_2$ and the associated production of HI. We define the basic physical quantities and dimensionless parameters in the problem, $\alpha$, $\beta$, ${\tilde \sigma}_g$, and $G$. We derive analytic expressions for the growth of the HI column density, from the outer PDR into the shielded cosmic-ray zone (CRZ), as a function of the gas density, far-UV field intensity, cosmic-ray ionization rate, and dust-to-gas ratio. In \S 2.3 we develop a formula for the critical cloud depths at which cosmic-rays dominate the the HI columns. In \S 2.4 we apply our formula to Galactic giant molecular clouds (GMCs) to assess whether cosmic rays can be significant contributors to HI columns in GMCs including their PDRs. In \S 3 we present numerical computations for the HI and H$_2$ abundance profiles and HI columns densities for a wide range of parameter combinations of the FUV intensity, cosmic-ray ionization rate, and gas density. We present results with and without the inclusion of a model for attenuation of the cosmic-ray fluxes. We also show how the profiles scale with the assumed dust-to-gas ratio. We discuss the effects of cosmic-ray ionization on the locations of the HI-to-H$_2$ transition points in the Appendix. We summarize in \S 4.

\section{Theory}

\subsection{Formation-Destruction Equation}

We consider an idealized one-dimensional semi-infinite cloud in slab geometry exposed on one side to beamed (normally incident) far-ultraviolet radiation, in combination with a flux of penetrating cosmic ray particles. In dusty systems the formation-destruction equation for the steady-state HI and H$_2$ fractions at any cloud depth is
\begin{equation}
    Rnx_{\rm HI} \ = \ [\frac{1}{2}D_0f(N_{\rm H_2})e^{-\tau_{\rm dust}} \ + \ {\phi}\zeta \frac{s(N)}{C}]x_{\rm H_2}
	\label{eq:formdes}
\end{equation}
where $x_{\rm HI}\equiv n_{\rm 
HI}/n$ is the atomic (HI) fraction, $x_{\rm H_2}\equiv n_{\rm H_2}/n$ is the molecular (H$_2$) fraction,
and $n_{\rm HI}$, $n_{\rm H_2}$, and $n$, are the atomic, molecular, and total hydrogen gas densities (cm$^{-3}$). Particle conservation is
\begin{equation}
    x_{\rm HI}+2x_{\rm H_2}=1 \ \ \ ,
\end{equation}
where we assume that the abundances of hydrogen species other than HI or H$_2$ are negligibly small\footnote{ See for example Fig.~10 in \cite{Bialy2015} or Fig.~6 in \cite{Sternberg2021}.}. 

The lefthand side of Eq.~(\ref{eq:formdes}) is the H$_2$ formation rate (s$^{-1}$), where
\begin{equation}
    R \equiv 3\times 10^{-17}\ {\tilde \sigma}_g \ T_2^{1/2} \ \ \ \ \ {\rm cm}^3\ {\rm s}^{-1}
    \label{eq:Rform}
\end{equation}
is the grain-surface H$_2$ formation rate coefficient, $T_2\equiv T/(100$~K) where $T$ is the gas temperature (K), and ${\tilde \sigma}_g$ is the dust-to-gas ratio normalized to  the standard Galactic ISM dust-to-gas mass ratio of 1:100 for which ${\tilde \sigma}_g$=1 \citep{Bohlin1978,RemyRuyer2014}.

The righthand side of Eq.~(\ref{eq:formdes}), is the H$_2$ destruction rate by Lyman-Werner band photodissociation (LW: 912-1108 \AA) in the PDR, and by cosmic-ray impact in the CRZ. In the first term,
\begin{equation*}
    D_0 \equiv 5.8\times 10^{-11}\ I_{\rm UV} \ \ \ \ \ {\rm s}^{-1}
\end{equation*}
is the unattenuated free-space rate \citep{Sternberg2014,Heays2017} for LW band photodissociation, 
\begin{equation}
   {\rm H_2 \ + \ \nu_{\rm LW} \ \rightarrow \ H \ + \ H} \ \ \ ,
\end{equation}
where $I_{\rm UV}$ is the far ultraviolet (6-13.6 eV) intensity relative to the \cite{Draine1978} representation for the interstellar radiation field in the Solar neighborhood ($I_{\rm UV}=1$). The factor of 1/2 accounts for the reduction of the photodissociation rate at the cloud surface due to the presence of the optically thick slab itself. The FUV and photodissociation rate are attenuated by a combination of dust absorption, and H$_2$ self-shielding as the LW absorption lines become optically thick.

The exponential term in Eq.~(\ref{eq:formdes}) accounts for the dust attenuation. The LW band dust optical depth
\begin{equation}
    \tau_{\rm dust}\equiv\sigma_g(N_{\rm HI} +2N_{\rm H_2})
    \label{eq:taudust}
\end{equation}
where $N=N_{\rm HI}+2N_{\rm H_2}$ is the total (atomic plus molecular) hydrogen column density from the cloud surface,
and
\begin{equation}
    \sigma_g\equiv 1.9\times 10^{-21} \ {\tilde \sigma}_g \ \ \ \ \ {\rm cm}^2 
    \label{eq:sigma}
\end{equation}
is the dust absorption cross section per hydrogen nucleus. Here ${\tilde \sigma}_g$ is the same dust-to-gas ratio appearing in Eq.~(\ref{eq:Rform}). This parameter can also be viewed as the normalized dust absorption cross section. I.e., we are assuming that the H$_2$ formation rate coefficient (Eq.~[\ref{eq:Rform}]) and the dust absorption cross section (Eq.~[\ref{eq:sigma}]) scale identically with the overall dust abundance. 

The H$_2$ self-shielding function
\begin{equation}
    f(N_{\rm H_2})\equiv \frac{1}{\sigma _d}\frac{dW_d(N_{\rm H_2})}{dN_{\rm H_2}}
\end{equation}
where $W_d(N_{\rm H_2})$ (Hz) is the multi-line curve of growth for the H$_2$ dissociation bandwidth, and $\sigma_d=2.36\times 10^{-3}$~cm$^{2}$~Hz is the total H$_2$ dissociation cross section (see \citetalias{Sternberg2014} for a detailed discussion of these quantities). We use the \cite{Draine1996} formula, as verified by \citetalias{Sternberg2014},  for the self-shielding function. 
At the cloud surface, $N_{\rm H_2}=0$ and $f=1$. For $N_{\rm H_2} \gtrsim 10^{14}$~cm$^{-2}$, the Doppler cores become optically thick and $f$ becomes small. For $N_{\rm H_2}\gtrsim 10^{22}$~cm$^{-2}$ the Lorentzian wings of the LW absorption lines overlap, and$f\rightarrow 0$.

In the second term in Eq.~(\ref{eq:formdes}), $\phi \zeta s(N)$ is the local destruction rate of the H$_2$ by the cosmic rays. Here,
\begin{equation}
    \zeta = 1.0\times 10^{-16} \ \zeta_{-16} \ \ \ \ \ {\rm s}^{-1}
\end{equation}
is the unattenuated free-space rate of H$_2$ ionization by cosmic-ray impact
\begin{equation}
    {\rm H_2 \ + \ cr \ \rightarrow \ H_2^+ \ + \ e} \ \ \ .
\end{equation}
This includes ionization by the primary cosmic-rays and the secondary energetic electrons.
The parameter $\phi$, of order unity, is the number of H$_2$ destruction events per cosmic-ray ionization. The H$_2$ destruction processes include ion-molecule chemical reactions driven by the initiating cosmic-ray ionizations, as well as direct cosmic-ray dissociation of the H$_2$. In a steady-state, the hydrogen gas is primarily a mixture of HI and H$_2$, and for a predominantly molecular medium $\phi\approx 2$  \citep{Bialy2015,Sternberg2021}. 

The factor $C$ in the second term accounts for possibly different H$_2$ formation rates in the CRZs compared to the PDRs due to density and temperature gradients, as well as additional gas clumping in the CRZs. Differing formation rates imply
\begin{equation}
    C \ = \ \frac{(nT^{1/2})_{_{\rm CRZ}}}{(nT^{1/2})_{_{\rm PDR}}} \ \ \ .
\end{equation}
For pressure equilibrium this then gives,
\begin{equation}
    C \ = \ \Bigl(\frac{n_{_{\rm CRZ}}}{n_{_{\rm PDR}}}\Bigr)^{1/2} \ = \ \Bigl(\frac{T_{_{\rm CRZ}}}{T_{_{\rm PDR}}}\Bigr)^{-1/2}  \ \ \ .
\end{equation}
For example, $C\approx \sqrt{5}$ for pressure equilibrium between a FUV heated PDR with $T\approx 100$~K, and a cosmic-ray heated CRZ with $T\approx 20$~K. $C$ can be increased further if there is any gas clumping.

The function $s(N)$ accounts for the possible attenuation of the cosmic-ray energy densities with cloud depth, and reduction of the associated H$_2$ ionization rates  \citep{Neufeld2017,Padovani2018,Sternberg2021}. However, the intrinsic energy spectra of the low-energy cosmic rays are uncertain, as are the transport mechanisms, e.g. free-streaming along magnetic field lines, or diffusive pitch-angle scattering off of pre-existing or self-generated MHD waves \citep{Zweibel2013,Padovani2020,Kempski2022}. In our computations we either exclude cosmic-ray attenuation entirely (and set $s=1$) or adopt a simple representative form for the cosmic-ray attenuation function. As in \cite{Sternberg2021}, when including cosmic-ray attenuation we adopt the \cite{Padovani2018} broken power-law model
\begin{equation}
s(N) = 
        \begin{cases}
      1 & \text{} \ N_{\rm eff} < N_{\rm cr} \\
       & \\
       (N_{\rm eff}/N_{\rm cr})^{-a}
      & \text{}\ N_{\rm eff} > N_{\rm cr} \ \ \ .
    \end{cases} 
    \label{eq:cosatt}
\end{equation}
Here $N_{\rm eff}\equiv N/{\rm cos}\theta$ is the effective absorbing gas column density, where $\theta$ is the angle of the magnetic field along which the cosmic-rays propagate relative to the cloud normal, and $N_{\rm cr}$ is the attenuation scale column. We use ``model ${\cal 
H}$" of \cite{Padovani2018} for which $a=0.385$, and $N_{\rm cr}=10^{19}$~cm$^{-2}$, and set ${\rm cos}\theta=1$. This model is in agreement with observed declines of the cosmic-ray ionization rates with increasing cloud column densities \citep{Caselli1998,Indriolo2012,Neufeld2017}. See Fig.~C1 in \cite{Padovani2022} for the full observational compilation. The power-law in Eq.~(\ref{eq:cosatt}) is valid for $N_{\rm eff}$ between $10^{19}$ and $10^{24}$ cm$^{-2}$. For $N_{\rm eff}<10^{19}$~cm$^{-2}$, $s=1$, and divergence is avoided at small columns.

Our basic question is: when (if ever) does internal cosmic-ray production of HI compete with photodissociation in the build-up of HI column densities in interstellar clouds? We are particularly interested in optically thick clouds consisting of fully developed outer photodissociation regions (PDRs) surrounding inner cosmic-ray dominated zones (CRZs).  We are interested in the total HI columns, irrespective of the temperature- and depth-dependent line widths of the associated 21 cm signatures.

\subsection{ODE}

\noindent
For our 1D geometry,
the atomic to molecular density ratio $x_{\rm HI}/x_{\rm H_2}\equiv dN_{\rm HI}/dN_{\rm H_2} $, and Eq.~(\ref{eq:formdes}) can be written as the ordinary differential equation (ODE)
\begin{equation}
    \frac{dN_{\rm HI}}{dN_{\rm H_2}} \ = \ \frac{1}{2}\alpha f(N_{\rm H_2})e^{-\sigma_g N} \ + \ \beta s(N) \ \ \ .
    \label{eq:formdes_D}
\end{equation}
In this equation the independent variable is $N_{\rm H_2}$ (with $N=N_{\rm HI}+2N_{\rm H_2}$) and the initial condition is $N_{\rm HI}(0)=0$. The parameters
\begin{equation}
    \alpha \equiv \frac{D_0}{Rn} \ = \ 1.9\times 10^4 \ \frac{I_{\rm UV}}{{\tilde \sigma}_gn_2}T_2^{-1/2} \ \ \ ,
    \label{eq:alpha}
\end{equation}
and
\begin{equation}
    \beta\equiv \frac{\phi\zeta}{RnC} \ = \ 6.7\times 10^{-2} \ \frac{\zeta_{-16}}{{\tilde \sigma}_gn_2C}T_2^{-1/2} \ \ \ ,
    \label{eq:beta}
\end{equation}
where $n_2\equiv n/(100$~cm$^{-3}$). Here and henceforth we assume $\phi=2$. The parameter $\alpha$ is the ratio of the unshielded free-space H$_2$ photodissociation rate to the molecular formation rate, and $\beta$ is the ratio of the unattenuated cosmic-ray destruction rate to the molecular formation rate. 

For characteristic interstellar conditions $I_{\rm UV}\approx 1$, and $\alpha \approx 1.9\times 10^4$ for a cold gas density $n_2=1$. The parameter $\alpha$ remains large even for $n_2 \gg 1$, especially near regions of active star-formation where the FUV field intensity $I_{\rm UV} \gg 1$.  At the cloud edge the H$_2$ is almost fully dissociated and the molecular fraction
\begin{equation}
    x_{\rm H_2} \ \approx \ \frac{2}{\alpha} \ = \ 
    1.1\times 10^{-4} {\tilde \sigma}_g T_2^{1/2} \frac{n_2}{I_{\rm UV}} \ \ \ .
\end{equation}
The molecular fraction grows as the FUV is attenuated with increasing cloud depth.

The Galactic cosmic-ray ionization rate also varies depending on location, with values approaching $10^{-15}$~s$^{-1}$ in diffuse gas down to $\sim 10^{-17}$~s$^{-1}$ in dense clouds \citep[e.g.][]{Caselli1998,Indriolo2012}. See also the observational summary in Fig.~C1 in \cite{Padovani2022}. Some of this variation may be indicative of attenuation of the cosmic-ray fluxes as they traverse the clouds \citep{Neufeld2017}. Here we adopt $\zeta_{-16}=1$ as a global characteristic value for the Galactic free-space cosmic-ray ionization rate.  With $\zeta_{-16}=n_2=T_2={\tilde \sigma}_g=1$, and with $C=\sqrt{5}$ for the CRZ, $\beta=3.0\times 10^{-2}$. The cosmic-ray ionization rates may be substantially larger in clouds near supernova remnants \citep{Indriolo2010,Ceccarelli2011,Schuppan2012}.

In most ISM environments $\alpha \gg 1$ and $\beta \lesssim 1$. For constant $Rn$ (independent of cloud depth) $\alpha$ is the maximal atomic to molecular density ratio at the fully photodissociated cloud edge, and $\beta$ is the minimal atomic to molecular ratio in the optically thick cosmic-ray dominated interior (in the absence of CR attenuation).  Complete atomic to molecular transitions are expected as the clouds become sufficiently optically thick. Given the solution to Eq.~(\ref{eq:formdes_D}) for $N_{\rm HI}(N_{\rm H_2})$, and with $N\equiv N_{\rm HI}(N_{\rm H_2})+2N_{\rm H_2}$ we obtain profiles for the HI column, $N_{\rm HI}$, and the derivatives, $x_{\rm HI}$ and $x_{\rm H_2}$, as functions of $N$. In \S 3 we present such profiles computed numerically, but we first discuss several analytic solutions, as follows.

\subsubsection{No CR attenuation}
In the absence of any CR attenuation, with $s\equiv 1$ everywhere, and for any $\beta$,
the HI fraction in the CRZ is,
\begin{equation}
    x_{\rm HI,CRZ} \ = \ \frac{\beta}{2 + \beta} \ \ \ .
    \label{eq:xHI}
\end{equation} 
The atomic fraction $x_{\rm HI}=1/2$ for $\beta=2$.
For a predominantly molecular CRZ, i.e.~for $\beta \ll 1$, the residual HI density is
\begin{equation}
    n_{\rm HI,CRZ} \ \approx \frac{\beta}{2}\ n_{\rm CRZ} \ = \ 3.33 \times  T_{\rm 2,CRZ}^{-1/2}\ {{\tilde \sigma}_g}^{-1}\zeta_{-16} \ \ \ \ \ \ {\rm cm}^{-3}
    \label{eq:nHI}
\end{equation}
{\it independent} of the total cloud density at any point (see also Solomon \& Werner 1971; Li \& Goldsmith 2003). For example, for $\zeta_{-16}={\tilde \sigma}_g=1$, and a CRZ temperature $T_{\rm 2,CRZ}=0.2$,  the HI density in the CRZ is $n_{\rm HI,CRZ}=7.4$~cm$^{-3}$.

For $\beta \lesssim 1$, and with $s\equiv 1$, an excellent approximate analytic solution to Eq.~(\ref{eq:formdes_D}) is
\begin{equation}
    N_{\rm HI}(N_{\rm H_2}) \ \approx \ \frac{1}{\sigma_g}{\rm ln}\bigl[\frac{\alpha}{2}G(N_{\rm H_2};\sigma_g) + 1\bigr] \ +  \ \beta N_{\rm H_2} \ \ \ .
    \label{eq:N1analytic}
\end{equation}
The first term on the right is the HI column built up by photodissociation, and the second term is the HI due to cosmic-rays, both as functions of the molecular column $N_{\rm H_2}$.
In this expression,
\begin{equation}
    \begin{split}
    G(N_{\rm H_2}; \sigma_g) \ \equiv \ & \sigma_g\int_0^{N_{\rm H_2}}f(N_{\rm H_2}^\prime)\ {\rm e}^{-2\sigma_gN_{{\rm H_2}}^\prime}\ dN_{\rm H_2}^\prime \\
    & = \ \frac{\sigma_g}{\sigma_d}W_{g}(N_{\rm H_2}; \sigma_g)
    \label{eq:GN2}
    \end{split}
\end{equation}
where $W_{g}(N_{\rm H_2}; \sigma_g)$ is the (universal) H$_2$-dust limited curve of growth for the LW dissociation bandwidth (see \citetalias{Sternberg2014} for a detailed discussion). For any $\sigma_g$, $W_g$ is a preexisting function of $N_{\rm H_2}$, independent of the cloud parameters $I_{\rm UV}$ or $n$. We use the analytic form for $W_g$ given by \citetalias{Bialy2016} (their Eq.~[27]). When all of the LW radiation is absorbed 
the integral converges to a constant, 
\begin{equation}
    G \ \equiv \ \frac{\sigma_g}{\sigma_d}W_{g,{\rm tot}}(\sigma_g) \ \approx \ 
    3.0\times 10^{-5}\ {\tilde \sigma}_g \biggl(\frac{9.9}{1+8.9{\tilde \sigma_g}}\biggr)^{0.37} \ \ \ .
    \label{eq:G}
\end{equation}
Here, $W_{g,{\rm tot}}(\sigma_g)$ is the total dust-limited dissociation bandwidth (Hz), and
$G$ is then the (dimensionless) average H$_2$ self-shielding factor within an H$_2$-dust absorption column. The righthand side of Eq.~(\ref{eq:G}) is our \citetalias{Bialy2016} fitting formula for $G$ based on the multi-line ($\it Meudon$) PDR model computations we presented in \citetalias{Sternberg2014}.

At cloud depths beyond which all of the LW radiation is absorbed, Eq.~(\ref{eq:N1analytic}) becomes
\begin{equation}
    N_{{\rm HI},t} \ \approx \ \frac{1}{\sigma_g}{\rm ln}\bigl[\frac{\alpha G}{2}+ 1\bigr] \ +  \ \frac{\beta}{2+\beta} N \ \ \ ,
    \label{eq:N1tot}
\end{equation} 
where the basic dimensionless parameter
\begin{equation}
    \alpha G \ = \ \frac{DG}{Rn} \ = \ 
    0.59\ \frac{I_{\rm UV}}{n_2}T_2^{-1/2}\times \biggl(\frac{9.9}{1+8.9{\tilde \sigma}_g}\biggr)^{0.37} \ \ \ .
    \label{eq:aG}
\end{equation}
The subscript $t$ refers to optically thick. The first term in Eq.~(\ref{eq:N1tot}) is the total (asymptotic) HI column density produced by just photodissociation in optically thick PDRs, 
\begin{equation}
    N_{\rm HI,PDR} \ \equiv \ \frac{1}{\sigma_g}{\rm ln}\Bigr[\frac{\alpha G}{2}+1\Bigr] \ \ \ .
    \label{eq:HItotFUV}
\end{equation}
This is the formula for the total HI column density for beamed FUV fields derived by \citetalias{Sternberg2014} in the absence of cosmic rays (i.e., for $\beta=0$). The second term in Eq.~(\ref{eq:N1tot})
\begin{equation}
    N_{\rm HI,CRZ} \ \equiv \ \frac{\beta}{2+\beta} N
        \label{eq:HItotCR}
\end{equation} 
is the additional HI column produced by the cosmic-rays, and it grows arbitrarily large with $N$ unless the cosmic-ray ionization rate is sufficiently attenuated. 
In this term we have used the relation $N_{\rm H_2}=N/(2+\beta)$ for the CRZ in replacing $N_{\rm H_2}$ with $N$ in Eq.~(\ref{eq:N1analytic}).

Differentiation\footnote{Differentiating Eq.~(\ref{eq:N1analytic}) gives $\tau_{\rm dust}=\sigma_g(N_{\rm HI}+(2-\beta)N_{\rm H_2})$, rather than Eq.~(\ref{eq:taudust}), for any $N_{\rm H_2}$. For $\beta<1$ the spurious term does not contribute significantly to the absorption of the FUV, and the shapes of the HI-to-H$_2$ profiles are unaffected.} shows that Eq.~(\ref{eq:N1analytic}) is a good (but formally approximate) solution to Eq.~(\ref{eq:formdes_D}) so long as $\beta\sigma_g N_{\rm H_2}$ is everywhere negligible compared to either $\sigma_g N_{\rm HI}$ or $2\sigma_g N_{\rm H_2}$. The latter two quantities are the HI-dust and H$_2$-dust optical depths associated with the HI and H$_2$ columns respectively. Thus, if $\beta=0$, i.e. with no cosmic-rays, Eqs.~(\ref{eq:N1analytic}) and (\ref{eq:N1tot}) are exact. 
But Eq.~(\ref{eq:N1analytic}) remains accurate, so long as $\beta < 1$.  We verify this in \S 3 by integrating Eq.~(\ref{eq:formdes_D}) numerically and comparing to our analytic formulae.

\subsubsection{With CR attenuation}

When CR attenuation is included, the HI fraction in the CRZ decreases with cloud depth as
\begin{equation}
    x_{\rm HI,CRZ} \ = \ \frac{\beta s(N)}{2+\beta s(N)} \ \approx \ \frac{1}{2}\beta s(N) \ \ \ .
    \label{eq:xHIwith}
\end{equation}
CR attenuation reduces the HI that is built up in the CRZs.

Our analytic approximation for $N_{{\rm HI},t}$ can be generalized for arbitrary cosmic-ray attenuation functions, $s(N)$, by making the replacement

\begin{equation}
    \begin{split}
    \frac{\beta}{2+\beta} N \ \rightarrow   \ \frac{\beta}{2+\beta}\int_0^N s(N^\prime)dN^\prime \\ 
     =  \frac{\beta}{2+\beta} \times 
        \begin{cases}
      N & \text{} \ N < N_{\rm cr} \\
       & \\ N_{\rm cr} +
      \frac{N_{\rm cr}}{1-a}\Bigl[\bigl(N/N_{\rm cr}\bigr)^{1-a}-1\Bigr] & \text{}\ N > N_{\rm cr}
    \end{cases} 
    \end{split}
    \label{eq:betaN2CR}
\end{equation}
for the second term in Eq.~(\ref{eq:N1tot}). In the second line we have evaluated the integral assuming the attenuation function given by Eq.~(\ref{eq:cosatt}) with ${\rm cos}\theta=1$.

\subsection{Critical Cloud Columns}
\label{sec:CritCols}

The two terms on the righthand side of 
Eq.~(\ref{eq:N1tot}) are equal at the critical gas column, $N=N_{\rm crit}$, at which the cosmic rays start to dominate the growth of the HI column. Thus, for unattenuated cosmic-rays ($s\equiv 1$),
\begin{equation}
    N_{\rm crit} \ = \ \frac{1}{\sigma_g} \frac{(2+\beta)}{\beta}\ {\rm ln}[\alpha G/2 + 1] \ \ \ .
    \label{eq:N_crit}
\end{equation}
Multiplying through by $\sigma_g$ gives the critical dust opacity,
\begin{equation}
        \tau_{\rm dust,crit} \ = \ \frac{(2+\beta)}{\beta}\ {\rm ln}[\alpha G/2 + 1] \ \ \ ,
    \label{eq:dust_crit}
\end{equation}
which depends on just the two (dimensionless) parameters $\alpha G$ and $\beta$. At a gas column $N$ (or dust opacity $\tau_{\rm dust}$), cosmic-rays dominate the production of the HI if $N>N_{\rm crit}$ (or if $\tau_{\rm dust}> \tau_{\rm dust,crit}$), otherwise photodissocation dominates.

In the weak-field limit, $\alpha G/2 \ll 1$ (and assuming $\beta \lesssim 1$), we have\footnote{For $G$ in the evaluation of Eq.(\ref{eq:N_cw}) we have dropped the term $[9.9/(1+8.9{\tilde \sigma}_g)]^{0.37}$, which varies by a factor of 4.2 for ${\tilde \sigma}_g$ between 0.1 and 10.}
\begin{equation}
        N_{{\rm crit},w} \ \approx \ \frac{1}{\sigma_g}\frac{\alpha G}{\beta} \ \approx \ 4.6\times 10^{21} \ \frac{I_{\rm UV}}{\zeta_{-16}} C \ \ \ \ \ {\rm cm}^{-2} \ .
    \label{eq:N_cw}
\end{equation}
 In this limit most of the HI produced by photodissociation is built up past the HI-to-H$_2$ transition point in gas that is primarily molecular (as shown in Fig.~\ref{fig:sig_1e0} in \S~\ref{sec:comps}, see also Appendix). When cosmic-rays are added the CRZs and PDRs overlap, and the critical column is a measure of the relative HI production efficiency by photodissociation versus cosmic-ray ionization in the molecular gas. The critical column is therefore proportional to $I_{\rm UV}/\zeta_{-16}$, independent of the cloud density $n$ and/or H$_2$ formation rate.

In the strong-field limit, $\alpha G \gg 1$ (and again for $\beta \lesssim 1$),
\begin{equation}
    N_{{\rm crit},s} \ \approx \ \frac{1}{\sigma_g}\frac{2\ {\rm ln}[\alpha G/2]}{\beta}  \ = \ 1.6\times 10^{22} \ \frac{n_2T_2^{1/2}C}{\zeta_{-16}}{\mathcal O}(1) \ \ {\rm cm}^{-2}.
    \label{eq:N_cs}
\end{equation}
For strong fields the photodissociated HI columns are built up in a (self-limited) fully atomic outer layer, and are only weakly (logarithmically) dependent on $I_{\rm UV}$. The cosmic-ray contributions to the HI occur in the inner fully optically thick regions where the atomic fractions are proportional to $\beta$. The critical gas column is therefore proportional to the ratio of the H$_2$ formation rate to ionization rate, or to the density to ionization rate for a given temperature, and the logarithmic factor of order unity. 

 In both the weak- and strong-field limits the critical columns are independent of the dust-to-gas ratio ${\tilde \sigma}_g$.

The intermediate case, $\alpha G/2\approx 1$, is also important, because for a narrow range around this UV to gas density ratio ($I_{\rm UV}/n_2\approx 3$) a two-phased (WNM/CNM) thermal equilibrium is possible for fully atomic (HI) gas \citep[\citetalias{Sternberg2014}]{Wolfire2003,Krumholz2008,Bialy2019}. The range for two-phased equilibria is $\alpha G \sim 1$ to 4, weakly dependent on metallicity. Star-forming gas and associated HI in the Milky Way and other galaxies may be self-regulated to be in a multiphased state \citep{Ostriker2010}. For such systems, the critical column is then
\begin{equation}
N_{\rm crit,CNM}\ = \ \frac{1}{\sigma_g}\frac{2\ {\rm ln(2)}}{\beta} \ = \ 1.1\times 10^{22} \ \frac{n_2T_2^{1/2}C}{\zeta_{-16}} \ \ \ \ \ {\rm cm}^{-2}
\label{eq:Nc2p}
\end{equation}
Here we are assuming that heating in the fully dissociated HI layers is via FUV photoelectric emission with negligible energy input by the cosmic-rays so that the thermal phase structure for the HI is primarily dependent on $\alpha G$, i.e. on the ratio $I_{\rm UV}/n$ \citep{Wolfire2003,Bialy2019}.

\subsection{Giant Molecular Clouds}
\label{sec:GMCs}

Our expressions for the critical cloud columns are for one-sided illumination by a  beamed (normally incident) FUV field, e.g.~for a cloud irradiated by a nearby hot star. In the ambient medium two-sided irradiation by the background FUV field is more appropriate, and the critical columns are then doubled for a given cosmic-ray ionization rate.
 Furthermore, the irradiation may be isotropic rather than beamed.

For example, for a Galactic giant molecular cloud (GMC) embedded within ambient photodissociated HI containing a two-phased mixture of CNM and WNM the two-sided intermediate case $\alpha G=2$ applies. Galactic GMCs have characteristic hydrogen column densities $N_{\rm GMC}\sim 1.5\times 10^{22}$~${\rm cm}^{-2}$ \citep{Solomon1987,McKee2007,Lada2020,Chevance2022} that are only weakly dependent on the cloud mass, $M_{\rm GMC}$, over a large range ($\sim 10$ to near $10^7$~M$_\odot$) implying a mass-radius relation that scales approximately as $M\sim R^2$ \citep{Larson1981}. 

With the extra factor of 2 for two-sided illumination,  and for beamed fields, it follows from Eq.~(\ref{eq:Nc2p}) that the GMCs are just critical for
\begin{equation}
    \beta \ = \ \frac{0.1}{{\tilde \sigma}_g} \times
 \Bigl(\frac{N_{\rm 22,GMC}}{1.5}\Bigr)^{-1} \ \ \ ,
    \label{eq:betacrit}
\end{equation}
or
\begin{equation}
    \frac{\zeta_{-16}}{n_2} \ = \ 1.5\ T_2^{1/2} \times C \ \Bigl(\frac{N_{\rm 22,GMC}}{1.5}\Bigl)^{-1} \ \ \ ,
    \label{eq:zetacrit_CNM}
\end{equation}
independent of ${\tilde \sigma}_g$.
Here $N_{\rm 22,GMC}\equiv N_{\rm GMC}/(10^{22} $cm$^{-2}$).
Thus, critical GMCs are molecular even without any CR attenuation, and for these $x_{\rm HI}\approx 0.05$.

For a spherical cloud, the average hydrogen density is
\begin{equation}
    {\bar n}_{\rm 2,GMC} \ = \ 0.83\ \Bigl(\frac{\langle N_{\rm 22,GMC}\rangle}{1.5}\Bigr)^{3/2}{M_{\rm 6,GMC}^{-1/2}} \ \ \ ,
    \label{eq:barn}
\end{equation}
where ${\bar n}_{\rm 2,GMC}\equiv {\bar n}_{\rm GMC}/(100\ $cm$^{-3}$), $\langle N_{\rm GMC}\rangle\equiv M_{\rm GMC}/\mu \pi R^2$  is the mean cloud column, and $M_{\rm 6,GMC}\equiv M_{\rm GMC}/(10^6$M$_\odot$). The mean mass per particle $\mu=2.34\times 10^{-24}$~g. 

 For spherical systems irradiation by ambient isotropic FUV fields is the more natural configuration \citep[e.g.,][]{McKee2010}. As discussed by \citetalias[]{Sternberg2014} the HI column produced by photodissociation on one side of a plane-parallel slab exposed to isotropic\footnote For a given H$_2$ photodissociation rate at the cloud surface, the incident radiation flux for isotropic fields is equal to half that for beamed fields, and $\alpha G$ is divided by 4 rather than 2 in Eq.~(\ref{eq:HItotFUViso}). The factor $\langle\mu\rangle=0.8$ is an angular average. See \citetalias{Sternberg2014} for a detailed discussion. radiation is
\begin{equation}
    N_{{\rm HI,PDR},i} \ = \ \frac{\langle\mu\rangle}{\sigma_g}{\rm ln}\Bigl[\frac{1}{\langle\mu\rangle}\frac{\alpha G}{4}+1\Bigr]
    \label{eq:HItotFUViso} \ \ \ ,
\end{equation}
where in this expression $\langle\mu\rangle = 0.8$. (The subscript $i$ refers to isotropic.) For a sphere, for which the PDR is a thin shell surrounding a CRZ core, the mean PDR HI column is
\begin{equation}
    \langle N_{\rm HI,PDR,i} \rangle \ = \ \frac{4\pi R^2\times N_{{\rm HI,PDR},i}}{\pi R^2} \ = \ 4\times N_{{\rm HI,PDR},i} \ \ \ .
\end{equation}
The mean CRZ HI column is
\begin{equation}
    \langle N_{\rm HI,CRZ} \rangle \ = \ \frac{\beta}{2+\beta} \ \langle N \rangle
\end{equation}
where $\langle N \rangle$ is the mean cloud column density. The critical mean column for a spherical cloud in an isotropic FUV field, and without CR attenuation, is then given by
\begin{equation}
    \langle N \rangle_{\rm crit} \ = \ \frac{2+\beta}{\beta} \times\frac{4\langle\mu\rangle}{\sigma_g}{\rm ln}\Bigl[\frac{1}{\langle\mu\rangle}\frac{\alpha G}{4}+1\Bigr] \ \ \ .
    \label{eq:Ncritsphere}
\end{equation}
This expression is analogous to Eq.~(\ref{eq:N_crit}) with the extra factor of 2 for two sided beamed illumination of a slab. The critical $\beta$ is hardly altered in switching from two-sided beamed slabs to isotropically illuminated spheres. For example, for $\alpha G=2$, the critical $\beta$ increases by just $\sim 10\%$ for spheres. Eqs.~(\ref{eq:betacrit}) and (\ref{eq:zetacrit_CNM})
are therefore unaltered for spheres, but with $N_{\rm 22,GMC}$ understood as the mean GMC column, as in Eq.~(\ref{eq:barn}).

Setting ${\bar n}_{\rm 2,GMC}T_{\rm 2,GMC}^{1/2}=n_2T_2^{1/2}C$ 
in Eq.~(\ref{eq:zetacrit_CNM}), 
and with Eq.~(\ref{eq:barn}), we obtain the critical mass 
\begin{equation}
    M_{\rm 6,GMC,crit} \ = \ 1.5 \ \frac{T_{\rm 2,GMC}}{\zeta_{-16}^2} \ \Bigl(\frac{N_{\rm 22,GMC}}{1.5}\Bigr) 
    \label{eq:Mcrit}
\end{equation}
below which the GMCs are FUV dominated, and above which they are cosmic-ray dominated.
With any gas clumping inside the GMC, or with the inclusion of CR attenuation, the critical masses will be larger still. 

It follows from Eq.~(\ref{eq:Mcrit}) that for a GMC temperature $T_{\rm 2,GMC}=0.2$ the critical ionization rate scales as
\begin{equation}
     \zeta_{\rm -16,crit} \ = \ 0.5 \times M_{\rm 6,GMC}^{-1/2}
 \ \ \ 
\end{equation}
for a standard $N_{\rm 22,GMC}=1.5$.
More massive GMCs require lower ionization rates to be cosmic-ray dominated because their mean densities are lower.
For example, for $6\times 10^6$~M$_\odot$ near to the upper end of the Galactic GMC mass distribution \citep{Williams1997} $\zeta_{\rm -16,crit}\approx 2\times 10^{-17}$~s$^{-1}$. 
For a more typical GMC mass of $10^4$~M$_\odot$, the critical ionization rate is $5\times 10^{-16}$~s$^{-1}$.

Relaxing the multiphase requirement, and deriving instead the critical column using Eq.~(\ref{eq:Ncritsphere}) for weak ($\alpha G \ll 1$)  isotropic FUV fields,  
GMCs are critical for
\begin{equation}
    \beta \ = \ \frac{7.0\times 10^{-2}}{{\tilde \sigma}_g} \times \Bigl(\frac{N_{\rm 22,GMC}}{1.5}\Bigr)^{-1} \ \alpha G \ \ \ ,
    \label{eq:weakcritaG}
\end{equation}
or for
\begin{equation}
    \zeta_{-16} \ = \ 0.6 \ \Bigl(\frac{N_{\rm 22,GMC}}{1.5}\Bigr)^{-1} \ I_{\rm UV} \ \ \ ,
        \label{eq:weakcritI}
\end{equation}
independent of ${\tilde \sigma}_g$, and independent of the gas density or cloud mass.  We stress again that Eqs.~(\ref{eq:weakcritaG}) and (\ref{eq:weakcritI}) hold for either slabs exposed to two-sided beamed fields or spheres illuminated by isotropic radiation. In Eq.~(\ref{eq:weakcritI}) we have assumed that $C=1$ since the PDRs and CRZs overlap in this limit (see \S~\ref{sec:HIH2profiles}).
Remarkably, in the weak-field limit, and for an ambient $I_{\rm UV}\approx 1$, the critical ionization rate for typical GMCs is close to the characteristic Galactic ionization rate $\zeta_{\rm -16} \approx 1$.

Our analysis of the GMCs thus far does not include the effects of CR attenuation, which we do consider in \S~\ref{sec:GMCsB} below. CR attenuation increases the critical ionization rates further.


\section{Computations}
\label{sec:comps}

\begin{figure}
    \label{fig:enter-label}
	\centering
\includegraphics[scale=0.5]{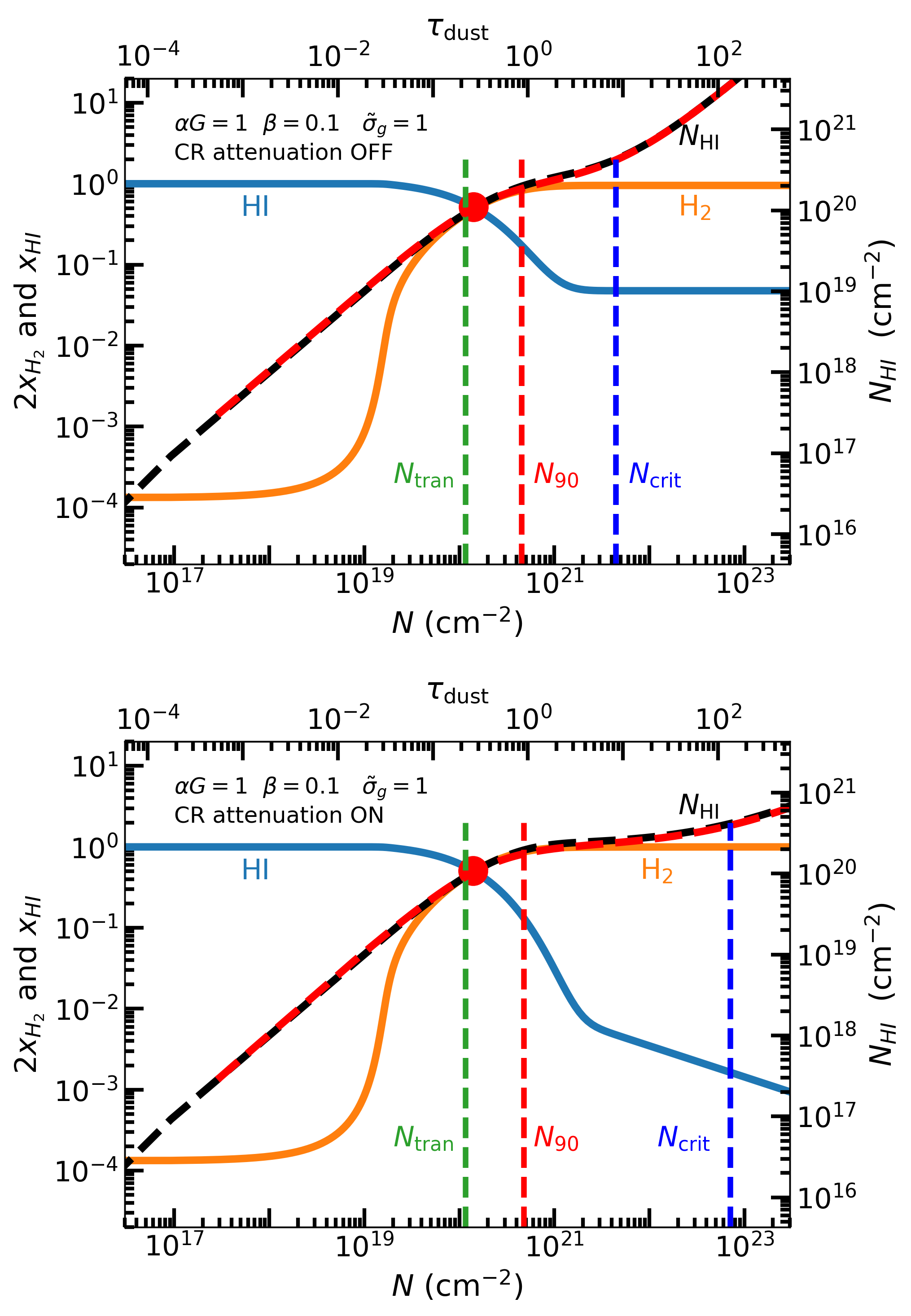}
	    \caption{HI and H$_2$ density profiles, for $\alpha G=1=0.59I_{\rm UV}/n_2$, with $T_2=1$ and ${\tilde \sigma}_g=1$ (see Eq.[\ref{eq:aG}]) and $\beta=0.1=3.0\times 10^{-2}\zeta_{-16}/{n_2}$, with $C=\sqrt{5}$  (see Eq.~[\ref{eq:beta}]). The upper panel are results without CR attenuation. The lower panel includes CR attenuation. The profiles are functions of the gas column $N$ (lower $x$-axes ) and the dust optical depth $\tau_{\rm dust}$ (upper $x$-axes). The curves are for the HI fractions $x_{\rm HI}$ (blue), twice the H$_2$ fractions $2x_{\rm H_2}$ (orange), and the HI column densities $N_{\rm HI}$ integrated numerically (dashed black), and using our analytic formula Eq.~(\ref{eq:N1analytic}) (dashed red). The left $y$-axes show the fractions, and the right $y$-axes are the column density scales.
      The red dots mark the HI-to-H$_2$ transition points where $x_{\rm HI}=2x_{\rm H_2}$. The analytic approximations for the transition points (Eq.[\ref{eq:tran}]) are indicated by the vertical dashed green lines. The vertical dashed red line is $N_{90}$ where 90\% of the photodissociated HI columns are built up.  The vertical dashed blue lines mark the critical cloud columns $N_{\rm crit}$ where the cosmic-ray contributions to the HI columns start to dominate.}
         \label{fig:fiducial}
\end{figure}

We now present numerical computations of the HI and H$_2$ density profiles and integrated HI column densities produced in gas slabs that are irradiated by combined fluxes of FUV photons and cosmic-rays. 
As is indicated by our analytic expressions Eqs.~(\ref{eq:N1analytic}) and (\ref{eq:N1tot}), the basic dimensionless parameter for the FUV driven HI-to-H$_2$ density profiles is $\alpha G$ (Eq.~[\ref{eq:aG}]) rather than $\alpha$ alone (see also \citetalias{Sternberg2014}). 

For the cosmic-rays the basic parameter is $\beta$ (Eq.[\ref{eq:beta}]). In this paper we are focussing on the regime $\beta \lesssim 1$ for which the gas is molecular in the absence of FUV, even without any CR attenuation. But the residual atomic component produced by the cosmic-ray bombardment contributes to the build up of the HI column densities.

We consider a wide range of conditions i.e., a range of $\alpha G$, and $\beta$, for varying dust-to-gas abundance ratios ${\tilde \sigma}_g$, and we present results with and without the inclusion of cosmic-ray attenuation. 
We use {\it Scipy} {\texttt {ODEINT}} to integrate Eq.~(\ref{eq:formdes_D}) and solve for $x_{\rm HI}/x_{\rm H_2}$ as a function of gas column $N\equiv N_{\rm HI}+2N_{\rm H_2}$, subject to $x_{\rm HI}+2x_{\rm H_2}=1$, for any $\alpha G$, $\beta$ and ${\tilde \sigma}_g$. When including CR attenuation we use the simple power-law form Eq.~(\ref{eq:cosatt}) for $s(N)$, assuming a normal magnetic field orientation ${\rm cos}\theta=1$ in the definition of $N_{\rm eff}$. We compare our numerical integrations to our analytic approximations for $N_{\rm HI}(N)$ given by Eqs.~(\ref{eq:N1analytic}) and (\ref{eq:N1tot}). 

As our first example, in Fig.~\ref{fig:fiducial} we show results for $\alpha G=1$, $\beta=0.1$, and ${\tilde \sigma}_g=1$. The FUV radiation and cosmic-rays are incident from the left (one-sided irradiation). In the upper panel, CR attenuation is not included. In the lower panel CR attenuation is included. For $\alpha G=1$, and with ${\tilde \sigma}_g=1$, the ratio of the (unattenuated) FUV field intensity to the gas density $I_{\rm UV}/n_2=1.7$, for a temperature $T_2=1$ (see Eq.~[\ref{eq:aG}]). The average self-shielding factor $G=3.0\times 10^{-5}$ (Eq.~[\ref{eq:G}]).
For $\beta=0.1$, the ratio of the cosmic-ray ionization rate to the gas density is $\zeta_{-16}/n_2=3.33$, for $C=\sqrt{5}$, and $\phi=2$ (see Eq.~[\ref{eq:beta}]).  

We plot $x_{\rm HI}$ and $2x_{\rm H_2}$ (blue and orange curves) as functions of the hydrogen column density $N$. The corresponding dust opacity, $\tau_{\rm dust}$, is shown along the auxiliary (upper) $x$-axes. The black dashed curves are the depth-dependent HI column densities found in our numerical integration of Eq.~(\ref{eq:formdes_D}). The column density scale is shown along the righthand auxiliary $y$-axis.
The overlying red dashed curves show the analytically computed HI columns. The agreement between the numerical and the analytically computed HI columns is excellent. 

As expected, the hydrogen is primarily atomic at the cloud edges, and the molecular fractions are very small, with $x_{\rm H_2}\approx x_{\rm H_2}/x_{\rm HI}= 2/\alpha = 6.2\times 10^{-5}$. 
Within the CRZs the gas is primarily molecular. In the absence of CR attenuation the HI fractions approach a cosmic-ray floor $x_{\rm HI}=\beta/2=5\times 10^{-2}$ (upper panel). 

The red marker dots indicate the numerically computed HI-to-H$_2$ transition points, defined as the cloud depths where $x_{\rm HI}=2x_{\rm H_2}$ (or $x_{\rm HI}=1/2$). For the models in Fig.~\ref{fig:fiducial} these occur at $N_{\rm tran}=1.4\times 10^{20}$~cm$^{-2}$, or $\tau_{\rm dust,tran}=0.26$. The vertical dashed green lines indicate these positions as given by the analytic \citetalias{Bialy2016} formula for the transition point (their Eq.~[39]). We discuss this formula in the Appendix
(Eq.~[\ref{eq:tran})]) and its continued range of applicability when cosmic-rays are included. 

The vertical red dashed lines mark the columns, $N_{\rm 90}$, where 90\% of the incident FUV radiation is absorbed. This occurs at $\tau_{\rm dust}\sim 1$, and defines the inner edge of the PDR. The HI column produced by photodissociation is $N_{\rm HI,PDR}=2.1\times 10^{20}$~cm$^{-2}$ (see Eq.~[\ref{eq:HItotFUV}]). The vertical blue dashed lines mark the critical columns, $N_{\rm crit}$ where the cosmic-rays start dominating the growth of the HI columns. Without CR attenuation this occurs at $N_{\rm crit}=4.4\times 10^{21}$~cm$^{-2}$. The corresponding critical dust opacity is $\tau_{\rm dust,crit}=8.4$.

When CR attenuation is included (lower panel) the atomic fraction falls below the $\beta=0.1$ cosmic-ray floor of $5\times 10^{-2}$ without attenuation. The ``knee" in the HI profile near $N=2\times 10^{21}$~cm$^{-2}$ is where the more slowly attenuating CR ionization processes take over from exponentially reduced photodissociation in producing the HI. The atomic fraction continues to decline at greater cloud depths as $\beta s(N)/2$, and becomes very small.  The much reduced HI abundance in the CRZ when CR attenuation is included moderates the growth of the HI column density (see Fig.\ref{fig:fiducial}) and the critical column is now $N_{\rm crit}=7.1\times 10^{22}$~cm$^{-2}$, or $\tau_{\rm dust,crit}=135.3$. When CR attenuation is included the CRZ must be 135 times larger than the PDR for cosmic-rays to contribute signifcantly to the production of the HI.

\subsection{Model Grid: No CR Attenuation, and  ${\tilde \sigma}_g=1$}

\subsubsection{HI and H$_2$ Profiles}

\label{sec:HIH2profiles}

\begin{figure*}
\centering
	\includegraphics[width=\textwidth]{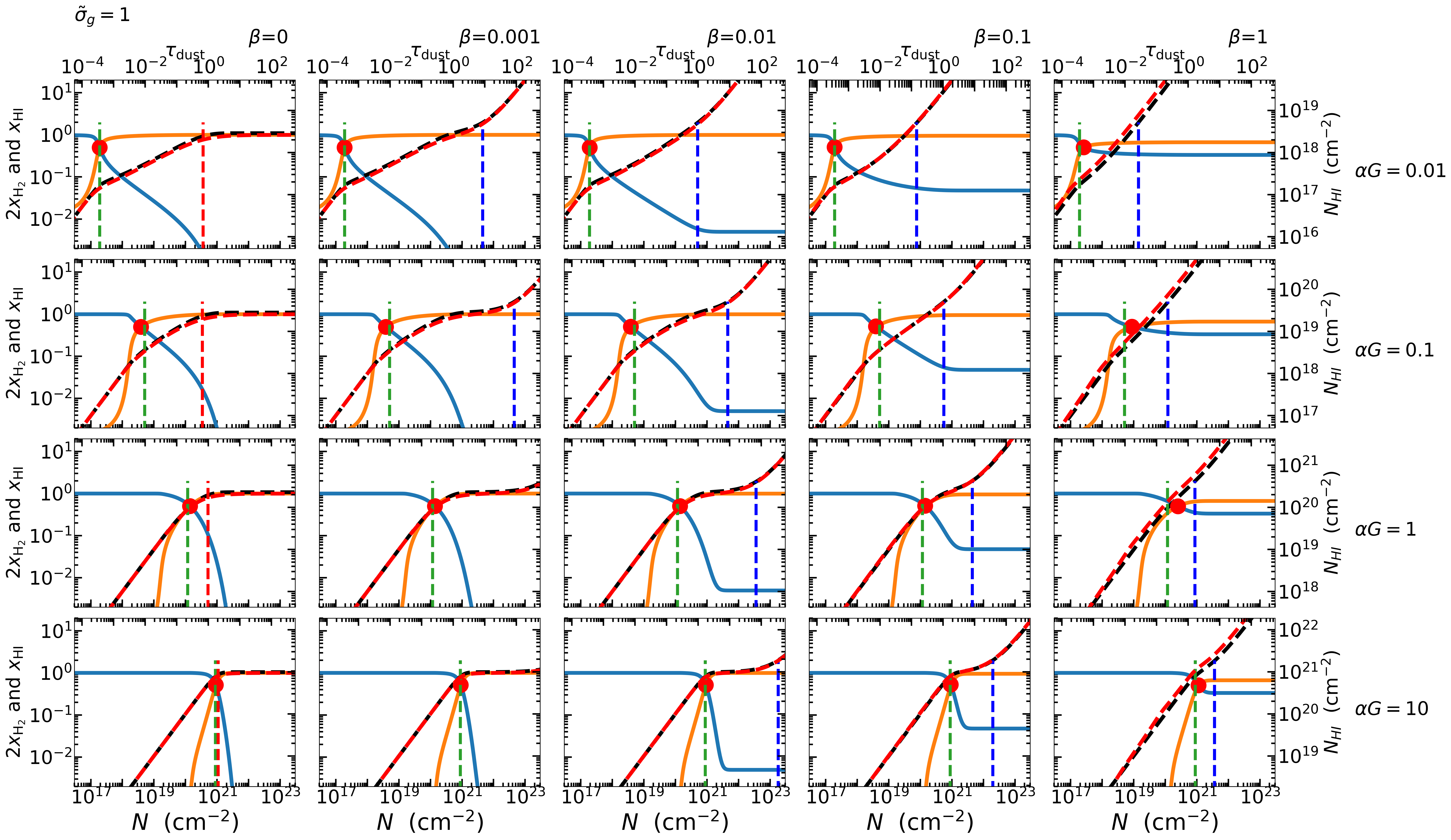}
	    \caption{ HI-to-H$_2$ density profiles as functions of the gas column $N$ (lower x-axes) and the dust optical depth $\tau_{\rm dust}$ (upper x-axes) for $\alpha G$ from 0.01 (weak field) to 10 (strong field), and for cosmic-ray parameters $\beta$ from 0 to 1, and with no cosmic-ray attenuation. The gas-to-dust ratio ${\tilde\sigma}_g=1$. The curves are for the HI fractions $x_{\rm HI}$ (blue), twice the H$_2$ fraction $2x_{\rm H_2}$ (orange), and the HI column density $N_{\rm HI}$, integrated numerically (dashed black), and using our analytic formula Eq.~(\ref{eq:N1analytic}) (dashed red). The red dots mark the HI-to-H$_2$ where $x_{\rm HI}=2x_{\rm H_2}$. The analytic approximation for the transition points (Eq.[\ref{eq:tran}]) are indicated by the vertical dashed green lines. For $\beta = 0$ (left column) the vertical dashed red lines are for $N_{90}$ where 90\% of the photodissociated HI columns are built up. For $\beta \neq 0$ the vertical dashed blue lines mark the critical cloud depths where the cosmic-ray contributions to the HI columns are equal to the photodissociated HI columns.}
    \label{fig:sig_1e0}
\end{figure*}

In Fig.~\ref{fig:sig_1e0} we present an $\alpha G$ versus $\beta$ model grid for the HI-to-H$_2$ density profiles, and integrated HI column densities, assuming ${\tilde \sigma}_g=1$, and no attenuation of the cosmic-ray ionization rates ($s=1$). From top to bottom, $\alpha G$ ranges from 0.01 (weak-field limit) to 10 (strong field limit). From left to right $\beta$ ranges from 0 to 1, i.e., weak to moderate\footnote{We reserve the term ``strong cosmic-ray irradiation" for $\beta > 2$ systems for which HI-to-H$_2$ transitions do not occur without CR attenuation. See Appendix.}
cosmic-ray irradiation. For ${\tilde \sigma}_g=1$ these ranges correspond to $I_{\rm UV}/n_2$ from $1.7\times 10^{-2}$ to 17.0, and $\zeta_{-16}/n_2$ from 0 to $33.3$ (for $C=\sqrt{5}$, $T_2=1$, and $\phi=2$).

As in Fig.~\ref{fig:fiducial},
in each panel we show the HI and H$_2$ fractions, $x_{\rm HI}$ (blue curves) and $2x_{\rm H_2}$ (orange curves), as computed by integrating Eq.~(\ref{eq:formdes_D}) numerically. The black and red dashed curves are the HI column densities found in our numerical integrations and using our analytic formulae respectively. The agreement between the two curves is excellent across the entire parameter space.

Again, the hydrogen is atomic at the cloud edges, and the molecular fractions are very small, with $x_{\rm H_2}\approx x_{\rm H_2}/x_{\rm HI}= 2/\alpha$ from $6.2\times 10^{-3}$ to $6.2\times 10^{-6}$. Within the CRZs the gas is (by assumption) primarily molecular, and the HI fractions approach $x_{\rm HI}=\beta/2$, i.e. range from $5\times 10^{-4}$ to $0.5$, from small to moderate cosmic-ray ionization rates.
The red dots are the HI-to-H$_2$ transition points. For the range of $\alpha G$ in Fig.~\ref{fig:sig_1e0} these occur at gas columns, $N_{\rm tran}$, equal to $1.9\times 10^{17}$, $3.9\times 10^{18}$, $1.4\times 10^{20}$, and $9.1\times 10^{20}$~cm$^{-2}$, corresponding to dust optical depths, $\tau_{\rm dust}$, equal to $3.6\times 10^{-4}$, $7.4\times 10^{-3}$, 0.26, and 1.7. The vertical green dashed lines show these positions using the \citetalias{Bialy2016} formula, Eq.~(\ref{eq:tran}). 
In the weak-field limit (small $\alpha G$) an HI-to-H$_2$ transition is induced by H$_2$ self-shielding at small cloud depths where $\tau_{\rm dust} \ll 1$, and dust attenuation is irrelevant for the transition point. Most of the photodissociated HI column density is  built up {\it inside} the predominantly molecular zone,
up to $\tau_{\rm dust}\approx 1$ where the FUV is finally fully absorbed. In the strong field limit (large $\alpha G$) the fully atomic layer becomes sufficiently large that the dust associated with this layer (the ``HI-dust") dominates the absorption of the FUV. The transition to H$_2$ is then very sharp, and most of the HI column is produced in the outer fully dissociated layer.  
Because we are assuming $\beta \le 1$ the cosmic-rays do not inhibit transitions to molecular gas as the clouds become optically thick to the FUV, and the transition points are unaffected (see Appendix).

The vertical red dashed lines shown for the $\beta=0$ cases (leftmost column in Fig.~\ref{fig:sig_1e0}) show the FUV ``absorption columns", $N_{90}$, where 90\% of the photodissociated HI columns are built up. The 90\% absorption depths occur at $\tau_{\rm dust}\approx 1$, independent of $\alpha G$, and are unaffected by the presence of cosmic-rays. We do not display the $N_{90}$ lines for the $\beta\ne 0$ panels. Instead, for $\beta\ne 0$ the vertical blue dashed lines indicate the critical gas columns, $N_{\rm crit}$, and dust opacities, $\tau_{\rm dust,crit}$, where the cosmic-ray and FUV contributions to the integrated HI columns are equal.

\subsubsection{Critical Dust Opacities and Gas Columns:\\ No CR Attenuation}

Without significant cosmic-ray attenuation the HI column densities diverge with increasing cloud gas column (see Eq.~[\ref{eq:N1tot}]). The critical gas columns indicated by the blue vertical lines are consistent with Eqs.~(\ref{eq:N_crit})-(\ref{eq:Nc2p}). For example, for $\alpha G=0.01$ and $\beta=0.001$,
$\tau_{\rm dust,crit}\approx\alpha G/\beta = 10$ and $N_{\rm crit}=5.3\times 10^{21}$~cm$^{-2}$ (see Eq.~[\ref{eq:N_cw}]). As $\beta$ is increased for $\alpha G=0.01$, the critical point moves inward and the PDRs and CRZs overlap, as seen in the top row of Fig.~\ref{fig:sig_1e0}, and as expected for the weak-field limit. As another example, and now for the strong field limit, for $\alpha G=10$ and $\beta=0.1$, $\tau_{\rm dust,crit}\approx 2{\rm ln}(\alpha G)/\beta = 32$, and $N_{\rm crit}=1.7\times 10^{22}$~cm$^{-2}$ (see Eq.~[\ref{eq:N_cs}]). As $\beta$ is increased in the strong field limit, $\tau_{\rm dust,crit}$ approaches the sharp HI-to-H$_2$ transition point, and the blue dashed lines approach the green dashed lines in Fig.~\ref{fig:sig_1e0}.


\begin{figure*}
\centering
	\includegraphics[width=\textwidth]{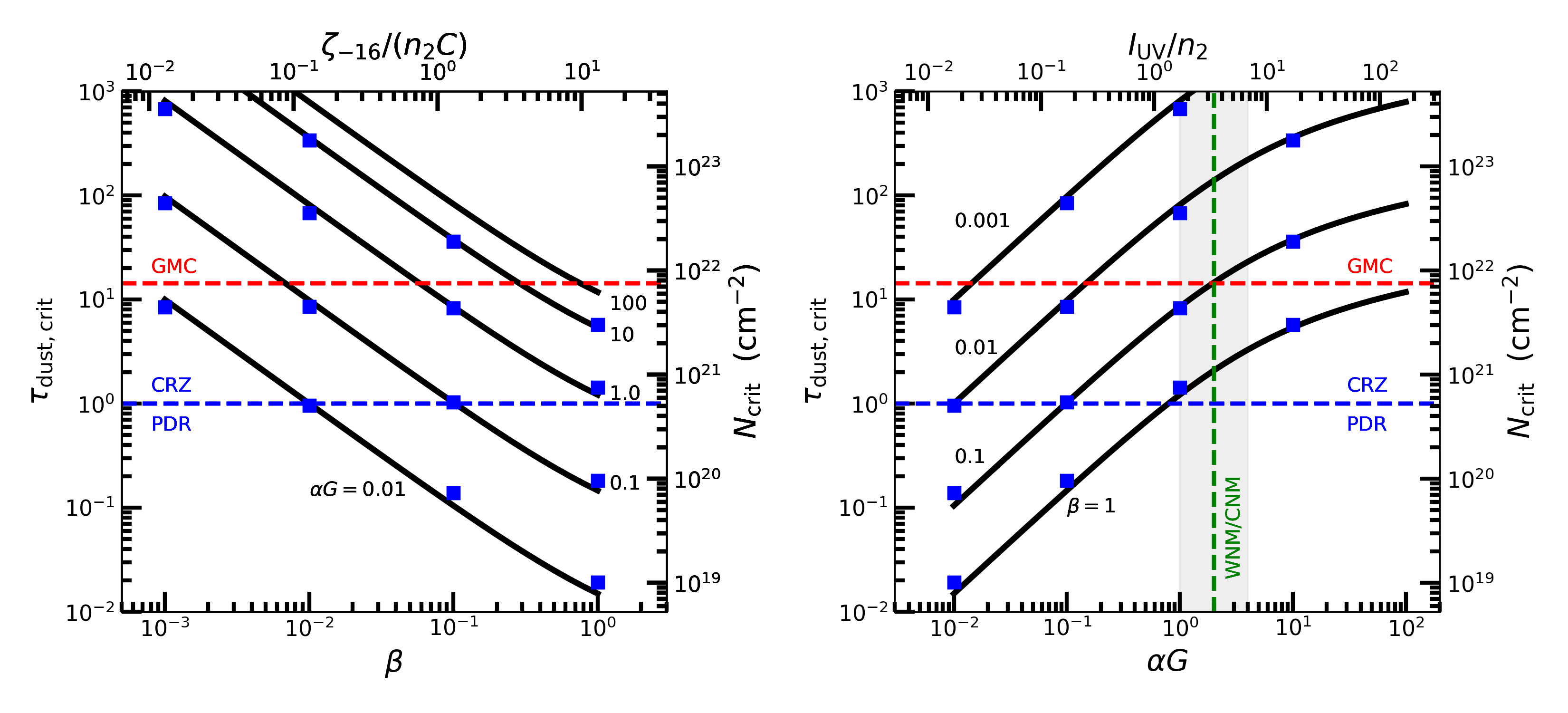}
	    \caption{Critical dust opacities, $\tau_{\rm dust,crit}$, at which the cosmic-ray contributions to the HI column densities are equal to the photodissociated HI columns. Right panel: $\tau_{\rm dust,crit}$ as functions of $\alpha G$ with individual curves as given by our analytic expression Eq.(\ref{eq:dust_crit}), for $\beta$ from 0.001 to 1. The blue squares are the numerically computed critical depths shown in Fig.~\ref{fig:sig_1e0}. The auxiliary axes for $I_{\rm UV}/n_2$ and $N_{\rm crit}$ are for a dust-to-gas ratio ${\tilde\sigma}_g=1$. The horizontal blue dotted line is for $\tau_{\rm dust}=1$ below which photodissociation and cosmic-ray production of the HI overlap (see text). The horizontal red dotted line corresponds to the typical half-column density of Galactic GMCs. The vertical green line marks the intermediate $\alpha G\approx 2$ regime for which multiphased HI is possible, and within the grey strip for which multi-phased HI is possible for FUV heated gas. Left panel: $\tau_{\rm dust,crit}$ as functions of $\beta$ with individual curves as given by our analytic expression Eq.(\ref{eq:dust_crit}), for $\alpha G$ from 0.01 to 100. The blue squares are the numerically computed critical depths shown in Fig.~\ref{fig:sig_1e0}. The auxiliary axes for $\zeta_{-16}/n_2C$ and $N_{\rm crit}$ are for a dust-to-gas ratio ${\tilde\sigma}_g=1$. The horizontal blue dotted line is for $\tau_{\rm dust}=1$ below which photodissociation and cosmic-ray ionization overlap. The horizontal red dotted line corresponds to the typical half-column density of Galactic GMCs.}
    \label{fig:taucrit}
\end{figure*}

In Fig.~\ref{fig:taucrit}, we plot curves as given analytically by Eq.~(\ref{eq:dust_crit}) for the critical dust opacities, $\tau_{\rm dust,crit}$, as functions of $\alpha G$ and $\beta$. The blue squares are
the critical opacities as found numerically in Fig.~\ref{fig:sig_1e0}. They lie very close to the analytic curves. The auxiliary $y$-axes in Fig.~\ref{fig:taucrit} show the corresponding critical gas column densities assuming ${\tilde \sigma}_g=1$ (Eq.~[\ref{eq:N_crit}]).  The left panel displays $\tau_{\rm dust,crit}$ as a function of $\beta$ (or $\zeta_{-16}/n_2C$ for ${\tilde \sigma}_g=1$) for several values of $\alpha G$ from 0.01 to 100. The right panel displays curves for $\tau_{\rm dust,crit}$ as a function of $\alpha G$ (or $I_{\rm UV}/n_2$ for ${\tilde \sigma}_g=1$) for several values of $\beta$ from 0.001 to 1. The curves illustrate the limiting behaviors given by Eqs.~(\ref{eq:N_cw}) and (\ref{eq:N_cs}). For a given $\alpha G$ the critical dust opacities and gas columns always vary inversely with $\beta$. For a given $\beta$, they vary linearly with $\alpha G$ in the weak-field limit ($\alpha G \ll 1)$ and logarithmically with $\alpha G$ in the strong-field limit ($\alpha G \gg 1)$. 

The horizontal dashed blue line in Fig.~\ref{fig:taucrit} is the $\tau_{\rm dust}=1$ boundary between the PDR and the CRZ. The curves again
show that in the weak-field limit, $\alpha G \ll 1$ cosmic-ray production of the HI can become competitive with photodissociation already within the PDRs (i.e.~within $\tau_{\rm dust}\lesssim 1$). 
Conversely, in the strong-field limit, $\alpha G \gg 1$, the critical opacities become large with $\tau_{\rm dust,crit} > 1$, even if $\beta$ approaches 1. In this limit the cosmic-ray production of the HI occurs mainly in the optically thick cloud interiors.

 
\subsection{Model Grid: With Cosmic-Ray Attenuation}

\subsubsection{HI and H$_2$ Profiles}

\begin{figure*}
\centering
	\includegraphics[width=\textwidth]{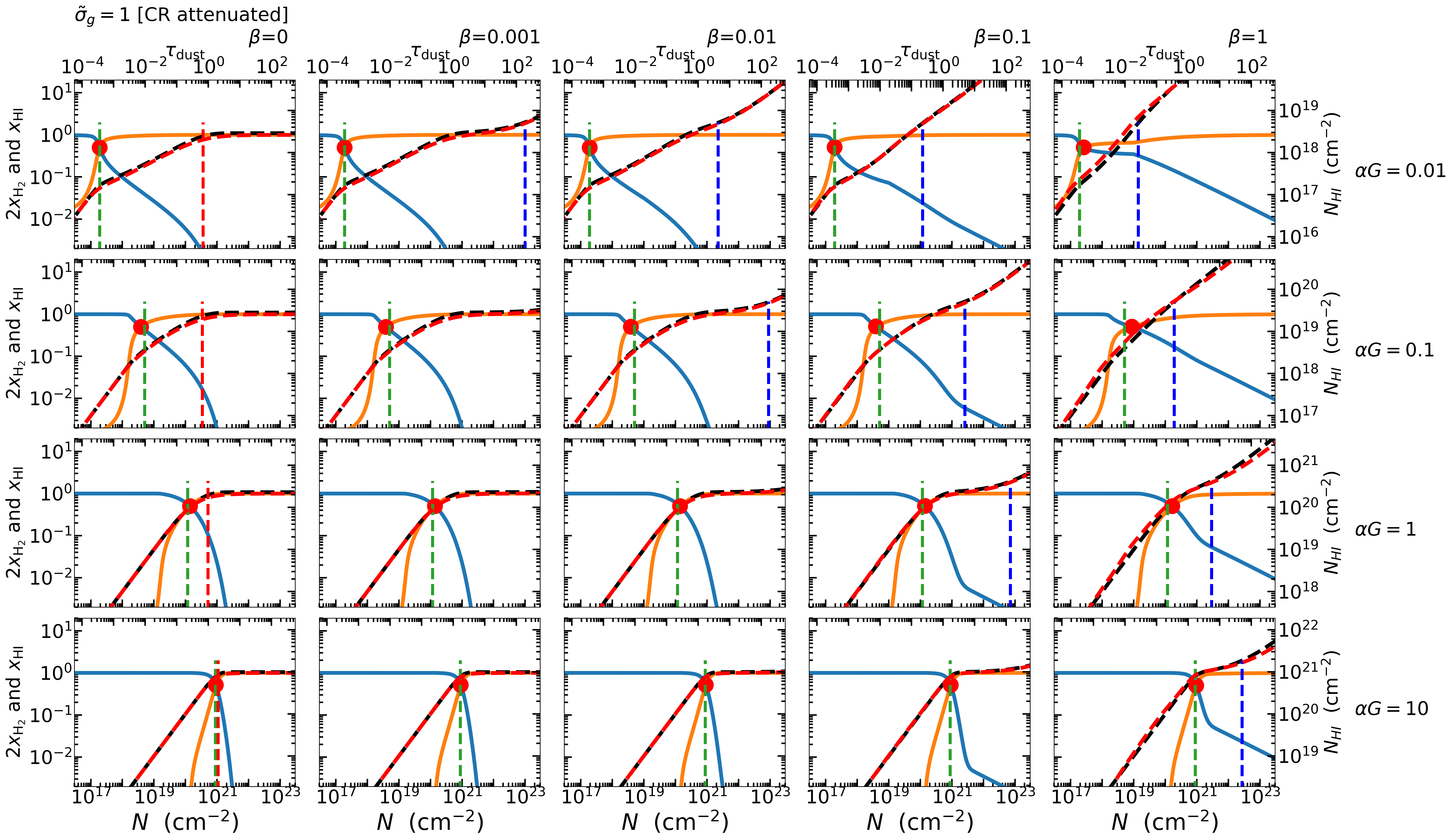}
	    \caption{Model grid with curves and markers as in Fig.~\ref{fig:sig_1e0}, but with the inclusion of CR attenuation.}
    \label{fig:sig_1e0_attenuated}
\end{figure*}

\begin{figure*}
\centering
	\includegraphics[width=\textwidth]{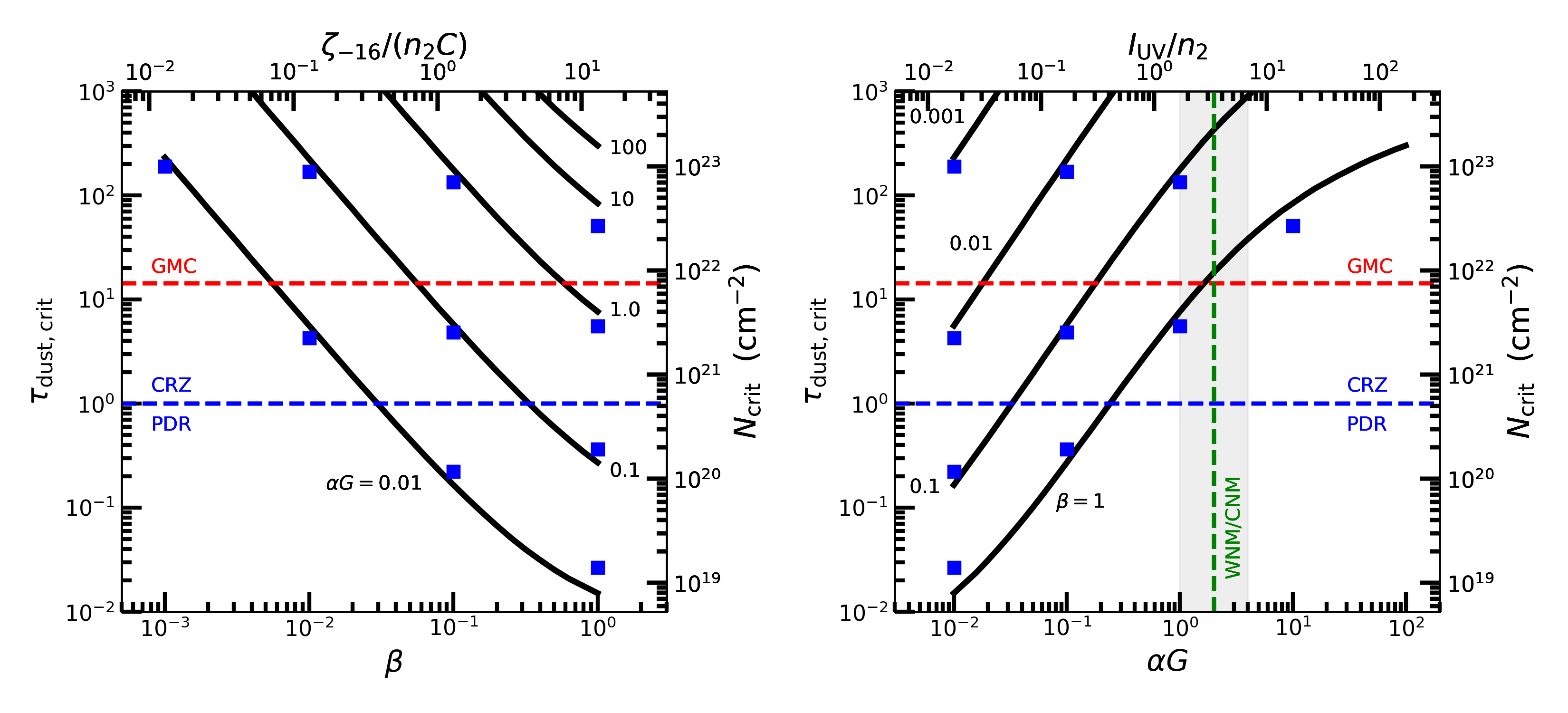}
	    \caption{As in Fig.~\ref{fig:taucrit} for the critical dust opacities, but with the inclusion of CR attenuation.}
    \label{fig:taucrit_attenuated.eps}
\end{figure*}

In Fig.~\ref{fig:sig_1e0_attenuated} we display the same $\alpha G$ versus $\beta$ grid for the HI and H$_2$ profiles as in Fig.~\ref{fig:sig_1e0} but now with the inclusion of cosmic-ray attenuation. We again assume ${\tilde \sigma}_g=1$.
In all panels, we assume the broken power-law CR attenuation function $s(N)$ as given by Eq.~(\ref{eq:cosatt}), with ${\rm cos}\theta=1$ for the magnetic field orientation. The effect of the cosmic-ray attenuation is most clearly seen for $\beta =1$ in the righthand column of Fig.~\ref{fig:sig_1e0_attenuated}. Without attenuation the HI fraction $x_{\rm HI}=1/3$ at large depths for $\beta=1$ (see Fig.~\ref{fig:sig_1e0}), and the integrated HI columns therefore rise sharply with increasing cloud depth. With attenuation the local HI fractions decrease and the resulting integrated HI columns are reduced.

The black dashed curves in Fig.~\ref{fig:sig_1e0_attenuated}  show the HI columns found by numerically integrating Eq.~(\ref{eq:formdes_D}) (again using {\it Scipy} {\texttt {ODEINT}}), but now including the attenuation function $s(N)$. The red dashed curves show the HI columns computed using our analytic approximation Eq.~(\ref{eq:N1analytic}) using Eq.~(\ref{eq:betaN2CR}) for the CR term. The agreement between the numerical solution and the analytic representation is excellent. 

As in Fig.~\ref{fig:sig_1e0} the vertical blue lines in Fig.~\ref{fig:sig_1e0_attenuated} mark the critical cloud depths at which the FUV and CR contributions to the HI columns are equal. Due to the reductions in the HI fractions in the CRZs the critical depths are increased compared to the no CR attenuation case. The effect is especially significant in the strong FUV field limit $\alpha G > 1$ for which the FUV contributions to the HI columns become large. (In some of the panels in Fig.~\ref{fig:sig_1e0} the blue markers do not appear because the critical depths are off scale high).


The red dots in Fig.~\ref{fig:sig_1e0_attenuated} show the HI-to-H$_2$ transition points where $x_{\rm HI}=2x_{\rm H_2}$. The vertical green lines mark the transition points as estimated using the \citetalias{Bialy2016} formula Eq.~(\ref{eq:tran}). The positions of the transition points are fully controlled by the FUV radiation absorption, and are not affected by the presence of cosmic rays or the inclusion of CR attenuation.

\subsubsection{Critical Dust Opacities and Gas Columns:\\ With CR Attenuation}

In Fig.~\ref{fig:taucrit_attenuated.eps}, we plot curves for the critical dust opacities, $\tau_{\rm dust,crit}$, as functions of $\alpha G$ and $\beta$, but now with the inclusion of CR attenuation as for the profiles shown in Fig.~\ref{fig:sig_1e0_attenuated}. To generate these curves we modify Eq.~(\ref{eq:dust_crit}) for $\tau_{\rm dust,crit}$ by making the replacement given by Eq.~(\ref{eq:betaN2CR}) in Eq.~(\ref{eq:N1tot}). The left panel displays curves for $\tau_{\rm dust,crit}$ as functions of $\beta$ (or $\zeta_{-16}/n_2C$ for ${\tilde \sigma}_g=1$) for several values of $\alpha G$ from 0.01 to 100. The right panel shows $\tau_{\rm dust,crit}$ versus $\alpha G$ (or $I_{\rm UV}/n_2$ for ${\tilde \sigma}_g=1$) for several values of $\beta$ from 0.001 to 1. The auxiliary $y$-axes in Fig.~\ref{fig:taucrit} show the corresponding critical gas column densities assuming ${\tilde \sigma}_g=1$.  
The blue squares are
the results of the numerical integrations found in Fig.~\ref{fig:sig_1e0_attenuated}, and they lie very close to the analytic curves. 
The primary affect of CR attenuation is to steepen the critical curves, since attenuation dampens the growth of the HI columns preferentially at low $\beta$ and large $\alpha G$.

For example, for $\alpha G=0.01$ and $\beta=0.001$,
$\tau_{\rm dust,crit}$ is increased to from 10 to 200 when CR attenuation is included, with
$N_{\rm crit}$ increasing to $1.1\times 10^{23}$~cm$^{-2}$. As $\beta$ is increased for $\alpha G=0.01$, the critical points move inward, the PDRs and CRZs overlap as seen in the top row of Fig.~\ref{fig:sig_1e0_attenuated}, and the attenuation effects are reduced due to the rapid build up of the CR contributions. As another example, for $\alpha G=1$ and $\beta=0.1$, $\tau_{\rm dust,crit}$ increases from 8.5 to 135. with $N_{\rm crit}$ increasing to $7.1\times 10^{22}$~cm$^{-2}$.

\subsection{GMCs and Multiphased HI}
\label{sec:GMCsB}

The horizontal red dashed lines in Figs.~\ref{fig:taucrit} and \ref{fig:taucrit_attenuated.eps} mark the half-column, $N_{\rm GMC}/2=7.5\times 10^{21}$~cm$^{-2}$, for typical Galactic GMCs, as discussed in \S~\ref{sec:GMCs}. For any $\alpha G$ and $\beta$ for which  $N_{\rm crit}<N_{\rm GMC}/2$, the GMC is ``supercritical" and the CRZ dominates the total HI column density. For $N_{\rm crit}>N_{\rm GMC}/2$ the GMCs are ``subcritical" and the PDR dominates the HI. GMCs are just critical for $\alpha G$ and $\beta$ at the intersections of the critical curves with the $N_{\rm GMC}/2$ line.

As discussed in \S~\ref{sec:GMCs},  without CR attenuation and in the weak-field limit GMCs are critical for $\beta\approx 7.0\times 10^{-2} \alpha G$, or for $\zeta_{\rm -16}\approx 0.6I_{\rm UV}$ (Eqs.~[\ref{eq:weakcritaG}] and [\ref{eq:weakcritI}]).  This relation is seen in Fig.~\ref{fig:taucrit} moving along the red line for small $\alpha G$. As shown in Fig.~\ref{fig:taucrit_attenuated.eps}, with CR attenuation the critical $\beta$ and $\zeta_{-16}$ are much larger. For example, for $\alpha G=0.1$, and without CR attenuation $\beta=7.0\times 10^{-3}$ for critical GMCs, and this increases to $6.0\times 10^{-2}$ when CR attenuation is included. Or, for $I_{\rm UV}=1$, the critical ionization rate $\zeta_{\rm -16}$ increases from 0.6 to 5.1 for models with and without CR attenuation respectively.

 The vertical green dashed lines in the righthand panels of Fig.~\ref{fig:taucrit} and \ref{fig:taucrit_attenuated.eps} 
 mark the intermediate $\alpha G= 2$ case (nominally $I_{\rm UV}/n_2\approx 3$), for which multiphased (WNM/CNM) HI is possible in the PDRs, as indicated by the grey strip. Without CR attenuation, and at $\alpha G=2$, the red GMC line in Fig.~\ref{fig:taucrit} intersects the critical curve for $\beta=0.1$. This is as given by Eq.~(\ref{eq:betacrit}).  Fig.~\ref{fig:taucrit_attenuated.eps} shows that $\beta=0.9$ when CR attenuation is included for critical GMCs with $\alpha G=2$. For example, for the nominal $I_{\rm UV}/n_2\approx 3$, the critical ratio $\zeta_{-16}/(n_2C)$ increases from 1.5 to 13.5. The critical free-space ionization rate then scales with GMC mass as 
\begin{equation}
\zeta_{\rm -16,crit}\approx 4.5\times M_{\rm 6,GMC}^{-1/2}
\end{equation}
when CR attenuation is included.

 As discussed in \S\ref{sec:GMCs}, the critical columns and ionization rates for spheres illuminated isotropically are essentially identical to the critical values for two-sided slabs, where the gas column $N$ for slabs is replaced by the mean column $\langle N \rangle$ for spheres. This is because $2N_{\rm HI,PDR,i}\approx N_{\rm HI,PDR}$ (see Eqs.~[\ref{eq:HItotFUV}] and [\ref{eq:HItotFUViso}]).

\begin{figure}
\centering
	\includegraphics[scale=0.5]{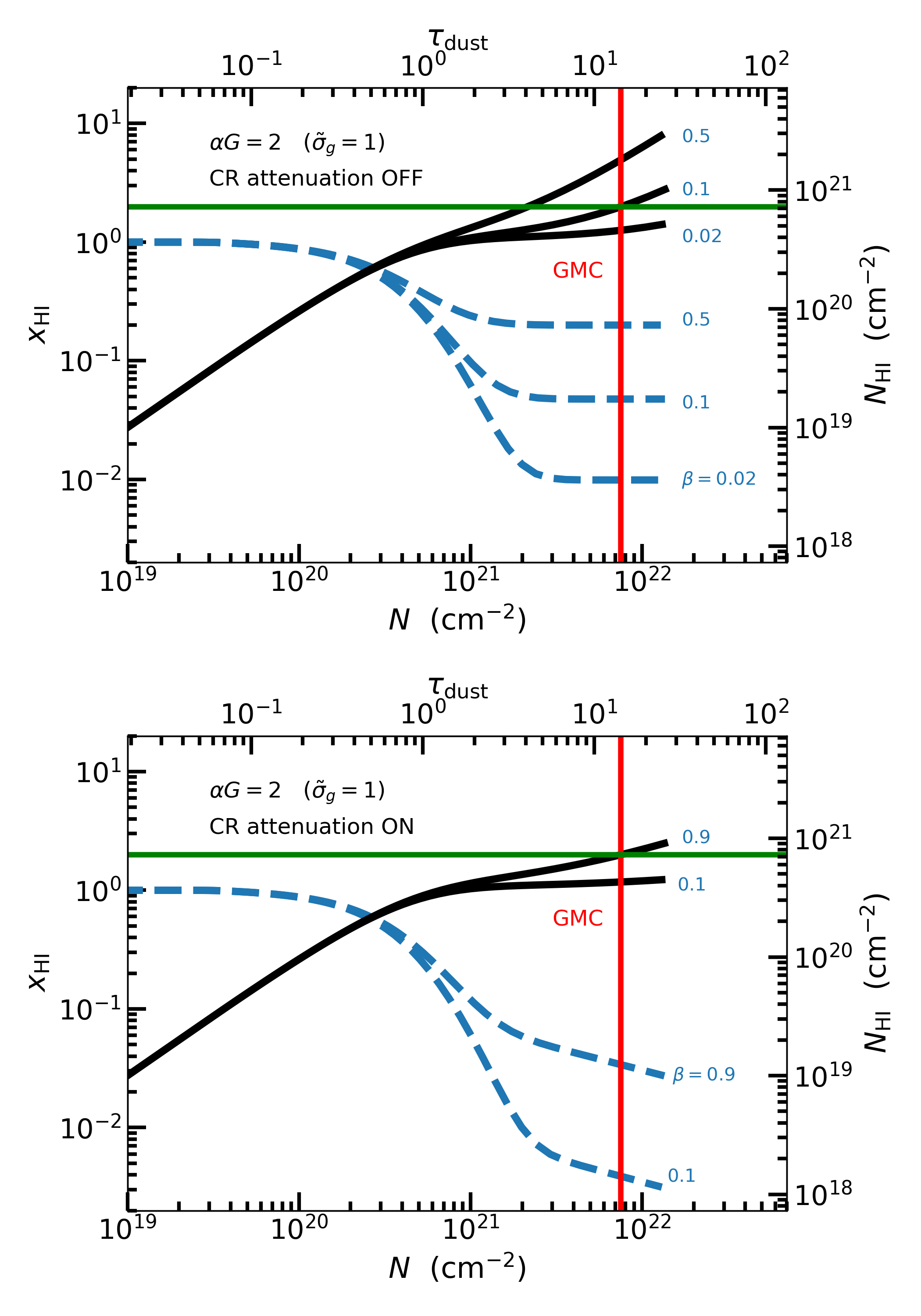}
	    \caption{HI fractions $x_{\rm HI}$ (dashed blue curves) and integrated HI columns $N_{\rm HI}$ (black curves) versus cloud depth, as parameterized by the gas column, $N$, or the dust optical depth, $\tau_{\rm dust}$, for the intermediate,  beamed field, $\alpha G=2$ case for multiphased HI, for a dust-to-gas ratio ${\tilde\sigma}_g=1$. The curves are labelled by the assumed values of $\beta$. In the upper panel CR attenuation is not included, and in the lower panel CR attenuation is included. The horizontal green line is for an HI column equal to twice the photodissociated column for $\alpha G=2$. The vertical red line marks the half gas column for typical Galactic GMCs.}
    \label{fig:GMC}
\end{figure}

 In Fig.~\ref{fig:GMC} we further consider the $\alpha G=2$ case. We show the HI fractions $x_{\rm HI}$ (blue dashed curves), and integrated HI column densities $N_{\rm HI}$ (black curves), for several values of $\beta$. 
 In the upper panel CR attenuation is excluded, and in the lower panel CR attenuation is included. For both the dust-to-gas ratio is ${\tilde \sigma}_g=1$. For $\alpha G=2$ the HI column produced by photodissociation is $N_{\rm HI,FUV}=3.65\times 10^{20}$~cm$^{-2}$. The horizontal green lines are at twice this value (see the righthand column density scale) and the intersections with the $N_{\rm HI}$ curves are at the critical cloud depths for each $\beta$. The vertical red lines mark the typical half-column of $7.5\times 10^{20}$~cm$^{-2}$ for the Galactic GMCs. Without CR attenuation we again see that GMCs are critical for $\beta=0.1$. With CR attenuation the critical value is much larger with $\beta=0.9$. This corresponds to a very large free-space ionization rate to density ratio $\zeta_{-16}/(n_2CT_2^{1/2})=13.4$.

\subsection{Dust-to-Gas Ratio}

How do the HI-to-H$_2$ profiles depend on the assumed dust-to-gas ratio, as parameterized by our ${\tilde\sigma}_g$? The dust-to-gas ratio (as controlled by the overall metallicity) enters in two ways. First via the H$_2$ dust-grain formation rate coefficient $R$ (eq.~[\ref{eq:Rform}]), and second via the FUV dust absorption cross section $\sigma_g$ (Eq.~[\ref{eq:sigma}]).
The formation rate coefficient, $R$, appears in the denominators of our dimensionless parameters $\alpha$ and $\beta$ (Eqs.~[\ref{eq:alpha}] and [\ref{eq:beta}]) in our ODE Eq.~(\ref{eq:formdes_D}). The dust absorption cross section, $\sigma_g$, appears in the definition of the dust optical depth, $\tau_{\rm dust}$, in Eq.~(\ref{eq:formdes_D}).  But importantly, the fundamental parameter $\alpha G$ is only weakly dependent\footnote{The term $(9.9/[1+8.9{\tilde\sigma}])^{0.37}$ in the definition of $G$ (Eq.~[\ref{eq:G}]) accounts for the dependence of the ``H$_2$-dust limited photodissociation rate" on ${\tilde\sigma}_g$. To keep $\alpha G$ fixed when varying ${\tilde\sigma}_g$ requires a corresponding alteration of $\alpha$ or the ratio $I_{\rm UV}/n$. See \citetalias{Sternberg2014} for a detailed discussion.} on ${\tilde\sigma}_g$ due to the cancellation when taking the ratio $\sigma_g/R$ (see Eq.~[\ref{eq:aG}]).

\citetalias{Bialy2016} studied the $\beta=0$ case (i.e. no cosmic rays) and found that when expressed in terms of $\tau_{\rm dust}$ (rather than the gas column $N$) then to a very good approximation, especially for ${\tilde\sigma}_g$ in the range 0.1 to 10, 
the HI-to-H$_2$ transition points depend on just $\alpha G$ independent of ${\tilde\sigma}_g$. This is the essence of our Eq.~(\ref{eq:tran}). This invariance is somewhat surprising especially in the weak-field limit where the FUV attenuation is governed purely by H$_2$ self-shielding, $\tau_{\rm dust,tran} \ll 1$, and dust-shielding plays no role. Furthermore, at cloud depths beyond the transition points, i.e.~within the molecular zones, the HI and H$_2$ density profiles as functions of $\tau_{\rm dust}$, also depend on just $\alpha G$, and are invariant with ${\tilde\sigma}_g$ to a very good approximation. Within the fully atomic outer layers, and up to the invariant transition points, the H$_2$ profiles do depend on ${\tilde\sigma}_g$, with H$_2$ fractions at the optically thin cloud edges that vary inversely  with the dust abundance and the associated H$_2$ formation rate coefficient.

\begin{figure}
\centering
	\includegraphics[scale=0.5]{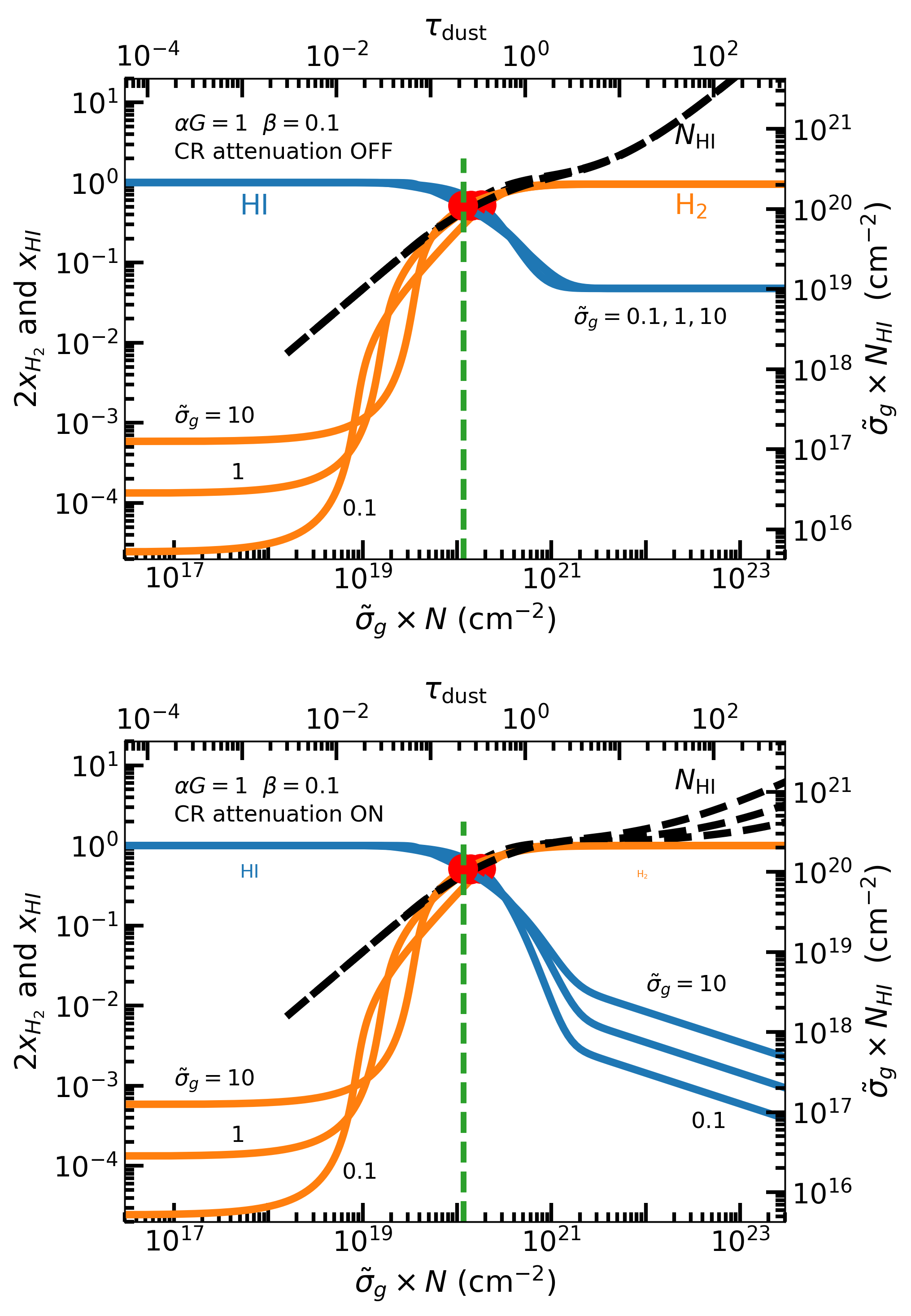}
	    \caption{HI and H$_2$ fractions, $x_{\rm HI}$ and $2x_{\rm H_2}$ (blue and orange curves) and HI column densities, $N_{\rm HI}$ (black curves) assuming $\alpha G=1$, and $\beta=0.1$, for ${\tilde \sigma}_g=$0.1, 1, and 10, without and with CR attenuation (upper and lower panels). The curves are plotted as functions of the dust optical depth $\tau_{\rm dust}$, or equivalently ${\tilde \sigma}_g\times N$, where $N$ is the gas column density.}
     \label{fig:sigma}
\end{figure}



Fig.~\ref{fig:sigma} illustrates the behavior when cosmic rays are included. In this example we set $\alpha G=1$ and $\beta=0.1$ and present results for ${\tilde \sigma}_g=0.1$, 1, and 10. In the upper panel we exclude CR attenuation, and in the lower panel CR is included according to Eq.~(\ref{eq:cosatt}). Once again, the blue and orange curves are the atomic and molecular fractions $x_{\rm HI}$ and $2x_{\rm H_2}$, and the black curves are the integrated HI column densities. The red markers indicate the transition points, as do the vertical green dashed lines according to Eq.~(\ref{eq:tran}).

For the three values of ${\tilde \sigma}_g$, the average H$_2$ self-shielding factor $G=5.5\times 10^{-6}$, $3\times 10^{-5}$, and $1.3\times 10^{-4}$, and $\alpha =1.8\times 10^5$, $3.3\times 10^4$ and $7.5\times 10^3$.
As seen in Fig.~\ref{fig:sigma} the corresponding molecular fractions at the cloud edges, $2x_{\rm H_2}=4/\alpha$, are $2.2\times 10^{-5}$, $1.2\times 10^{-4}$, and $5.3\times 10^{-4}$. When expressed in terms of the dust optical depth $\tau_{\rm dust}$ (or equivalently ${\tilde \sigma}_g\times N$) the transition point is insensitive to ${\tilde \sigma}_g$, and for our assumed $\alpha G$ the transitions occur at $\tau_{\rm dust}=0.2$. For our assumed $\beta$ the cosmic rays do not affect the positions of the transition points (see Appendix). 

The cosmic-ray ionization rate is unaffected by the presence of dust, and in the absence of CR attenuation the HI profiles within the optically thick CRZs do not depend on ${\tilde \sigma}_g$ and the atomic fraction reaches the cosmic-ray floor $x_{\rm HI}=4.8\times 10^{-2}$ (see upper panel). However, with CR attenuation, and when expressed as functions of the dust optical depth, the HI fractions increase with ${\tilde \sigma}_g$ for a given $\tau_{\rm dust}$ (see lower panel).  This is simply because the CR attenuation depends on the gas {\it column} $N=\tau_{\rm dust}/\sigma_g$. For our assumed CR attenuation power-law (Eq.~[\ref{eq:cosatt}]) with $a=0.385$, $x_{\rm HI}$ varies as ${\tilde \sigma}_g^{0.385}$ at a given dust optical depth. In all cases the atomic fractions decrease with cloud depth ([Eq.~\ref{eq:xHIwith}]).

We have verified by explicit computation that this ovearall behavior is maintained for the entire range of $\alpha G$ and $\beta$ in our model grids for ${\tilde \sigma}_g$ from 0.1 to 10.

\section{Summary}

In this paper we extend the analytic treatment presented by \cite{Sternberg2014} and \cite{Bialy2016} (\citetalias{Sternberg2014} and \citetalias{Bialy2016}) for the production of atomic hydrogen (HI) via FUV photodissociation at the boundaries of interstellar molecular (H$_2$) clouds, to also include the effects of penetrating (low-energy) cosmic-rays for the growth of the total HI column densities. 

We focus on idealized one-dimensional gas slabs, consisting of outer photodissociation regions (PDRs) and inner cosmic-ray zones (CRZs).  
We compute the depth dependent steady-state abundances of the HI and H$_2$, in a balance between grain-surface formation of the H$_2$ and destruction via FUV photodissociation and cosmic-ray ionization. The FUV photodissociation rates are reduced by (standard) H$_2$ self-shielding, and dust absorption. For the cosmic-rays we assume either constant overall ionization rates, or models that include depth-dependent attenuation of the cosmic-ray fluxes. 

The physical parameters in the problem are (a) the free-space intensity, $I_{\rm UV}$, of the FUV radiation and the associated H$_2$ photodissociation rate; (b) the free-space cosmic ray H$_2$ ionization rate, $\zeta$; (c) the density, $n$, of hydrogen nuclei, in atoms and molecules; (d) the H$_2$ formation rate coefficient $R$; (e) the FUV dust absorption cross section $\sigma_g$; (f) the gas temperature $T$; and (g) a density enhancement factor, $C$, for the cool CRZs relative to the warmer PDRs.  An additional (chemical) parameter is the number, $\phi$, of H$_2$ dissociations per cosmic-ray ionization event.

The governing HI/H$_2$ formation-destruction equation that we solve is Eq.~(\ref{eq:formdes}), or in differential form Eq.~(\ref{eq:formdes_D}). The solutions for the HI and H$_2$ density profiles and the integrated HI columns, depend primarily on the ratios $I_{\rm UV}/Rn$ and $\zeta/Rn$, as encapsulated in our dimensionless parameters $\alpha G$, and $\beta$ (Eqs.~[\ref{eq:alpha}], [\ref{eq:beta}] and [\ref{eq:aG}]). A third dimensionless parameter is the dust-to-gas ratio ${\tilde \sigma}_g$. It sets the magnitude of both the dust absorption cross section, and the molecular formation rate coefficient.

We solve Eq.~(\ref{eq:formdes_D}) numerically, and we also develop simple analytic formulae for the growth of the HI column density in terms of $\alpha G$ and $\beta$ (Eqs.~[\ref{eq:N1analytic}] and [\ref{eq:N1tot}]). Our analytic formulae provide an excellent match to the numerical integrations. Our focus is on conditions ($\beta \le 1$) for which the gas is primarily molecular in the optically thick cloud interiors. As we discuss in the Appendix, for these conditions cosmic-rays do not affect the locations of the HI-to-H$_2$ transition points. We consider both weak fields ($\alpha G \ll 1$) and strong fields ($\alpha G \gg 1$), and compute the critical cloud columns, $N_{\rm crit}$, at which cosmic-rays dominate the production of the total HI columns. We write down analytic expressions for the critical columns.  We also examine how the HI and H$_2$ profiles scale with the assumed dust-to-gas ratio.

As an example, we apply our theory to Galactic giant molecular clouds (GMCs), with typical hydrogen gas column densities $\sim 1.5\times 10^{22}$~cm$^{-2}$ (independent of mass).  For GMCs we consider both plane-parallel slabs exposed to beamed FUV fields, and spherical clouds illuminated by isotropic radiation. For weak FUV fields, for which $I_{\rm UV}/n \ll 3.4\times 10^{-2}$~cm$^3$, and with $I_{\rm UV}=1$, the CRZ dominates the production of the HI if the free-space $\zeta> 5.1\times 10^{-16}$~s$^{-1}$.
This estimate for the critical ionization rate includes cosmic-ray attenuation within the GMCs. For multiphased warm/cold HI within the PDRs, for which $I_{\rm UV}/n \approx 3.4\times 10^{-2}$~cm$^3$, the CRZ dominates the HI if $\zeta \gtrsim 4.5\times 10^{-16} \times (M_{\rm GMC}/10^6 \ M_{\odot})^{-1/2}$~s$^{-1}$, where $M_{\rm GMC}$ is the GMC mass. The very large critical ionization rates suggest that FUV photodissociation dominates the production of the HI in most Galactic GMCs.

\section*{Acknowledgements}

We thank David Neufeld, Chris McKee, Eve Ostriker, and Mark Wolfire for discussions.  We thank the referee for a careful reading of our manuscript and for helpful comments.
This work was supported by the German Science Foundation via DFG/DIP grant STE/ 1869-2 GE/ 625 17-1, by the Center for Computational Astrophysics (CCA) of the Flatiron Institute, and by the Mathematical and Physical Sciences (MPS) division of the Simons Foundation, USA.


\bibliographystyle{mnras}


\appendix
\section{HI-to-H$_2$ transition points}

We define the HI-to-H$_2$ transition points as the depths where $x_{\rm HI}=2x_{\rm H_2}$ (or $x_{\rm HI}=1/2$).
In the absence of cosmic rays \citetalias{Bialy2016} derived the 
fitting formula 
\begin{equation}
    N_{\rm tran,PDR} \ = \ \frac{q}{\sigma_g}\ {\rm ln}[(\frac{\alpha G}{2})^{1/q}+1]
    \label{eq:tran}
\end{equation}for the gas column densities at which the transitions occur for pure FUV irradiation. We refer to these as the PDR transition points.
We emphasize that $N_{\rm tran,PDR}$ is the {\it total} gas column density, atomic plus molecular, at the PDR transition point. The associated dust optical depth is $\tau_{\rm dust,tran}\equiv \sigma_gN_{\rm tran}$, and Eq.~(\ref{eq:tran}) gives this optical depth as a function of our basic parameter $\alpha G$. In this formula, $q$ is the power-law index\footnote{In \citetalias{Bialy2016} we used the letter $\beta$ for the index. We have altered our notation here to keep $\beta$ for the cosmic-ray parameter (Eq.[\ref{eq:beta}]) as in SGB20.} for a simple power-law approximation for the H$_2$ self-shielding function (see \citetalias{Bialy2016}). The recommended value is $q=0.7$.  

Importantly, Eq.~(\ref{eq:tran}) is valid in both the strong- or weak-field regimes, i.e., whether the transitions are controlled by dust absorption or molecular self--shielding.
For strong-fields, $\alpha G \gg 1$,
\begin{equation}
\begin{split}
    N_{\rm tran,PDR} \ & = \ N_{\rm HI,PDR} \ \approx \ \frac{1}{\sigma_g}{\rm ln}[\alpha G/2] \\
    & = \ \frac{5.3\times 10^{20}}{{\tilde \sigma}_g} {\mathcal O}(1)\ \ \ {\rm cm}^{-2}
\end{split}
\end{equation}
 In this limit the total column density at the transition point is just the entire photodissociated HI column in the cloud as given by Eq.~(\ref{eq:HItotFUV}), with negligible H$_2$ up to this point. All of the HI is produced in an outer photodissociated layer, with a column of order $1/\sigma_g$, up to a sharp HI-to-H$_2$ transition. 
 
 In the weak-field limit Eq.~(\ref{eq:tran}) remains accurate, and without explicit reference to the H$_2$ column, even though the transition point is governed purely by H$_2$ self-shielding. In this limit it is straightforward to show that 
 \begin{equation}
     N_{\rm tran,PDR} \ \approx \frac{q}
{\sigma_g} (\alpha G/2)^{1/q} \approx \ (2-q)\times {N_{\rm HI,tran}} \ \ \ ,
 \end{equation}
 where $N_{\rm HI,tran}$ is the HI column density at the transition point. In this limit, $N_{\rm tran} \ll 1/\sigma_g$, and with $q=0.7$ we have 
 $N_{\rm HI,tran}\approx 0.77N_{\rm tran,PDR}$.

 When cosmic rays are included the transition points may be moved to greater depths than $N_{\rm tran,PDR}$ if the ionization rates are large enough \citep{Kim2023}. However, this requires extremely large ionization rates. 

  \begin{figure}
\centering
	\includegraphics[scale=0.5]{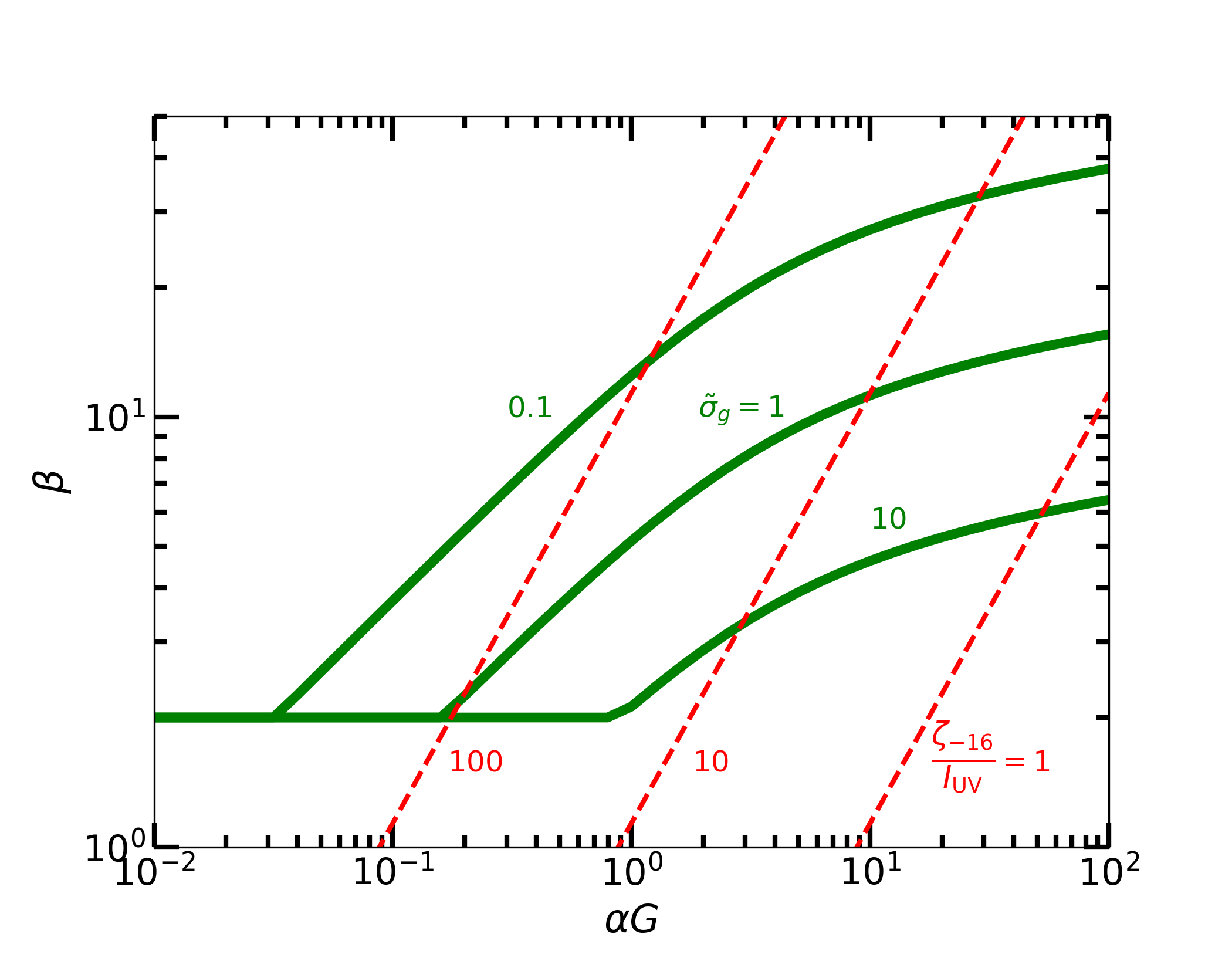}
	    \caption{Curves (in green) for $\beta_{\rm max}$ as given by Eq.~(\ref{eq:betaM}), for ${\tilde \sigma}_g$=0.1, 1, and 10. The lines (red dashed) are for constant ratios $\zeta_{-16}/I_{\rm UV}=$1, 10, and 100.}
    \label{fig:betaMAX}
\end{figure}
 
 The transition points are affected by cosmic rays if
 \begin{equation}
    \beta \ > \ \beta_{\rm max} \ \equiv \ \frac{2}{s(N_{\rm tran,PDR})} \ \ \ ,
    \label{eq:betaM}
 \end{equation}
 where $s(N) \le 1$ is the CR attenuation function. For these values of $\beta$ the cosmic ray ionization rates are large enough to keep $x_{\rm HI} > 1/2$ at the FUV only transition points. For $\beta < 2$ the transition points are unaffected and remain at $N_{\rm tran,PDR}$ as given by Eq.~(\ref{eq:tran}), whether or not CR attenuation is included.  Without CR attenuation ($s=1$) a transition never occurs if $\beta > 2$ since then the cosmic-rays maintain $x_{\rm HI} > 1/2$ everywhere. We refer to this as the ``strong cosmic-ray irradiation" limit. 
 With CR attenuation, the transition points are unaffected if $\beta < \beta_{\rm max}$.

 In Fig.~\ref{fig:betaMAX} we plot $\beta_{\rm max}$ versus $\alpha G$ using Eq.~(\ref{eq:tran}) for $N_{\rm tran,PDR}$, and Eq.~(\ref{eq:cosatt}) for $s(N)$. We show curves for ${\tilde \sigma}_g$ equal to 0.1, 1, and 10. For small $\alpha G$ the FUV transition points occur at $N_{\rm tran,PDR}< N_{\rm cr}=10^{19}$~cm$^{-2}$ for which $s=1$ (see Eq.[\ref{eq:cosatt}]) so that $\beta_{\rm max}=2$. As $\alpha G$ is increased the CR attenuation at the PDR transition points become significant and $\beta_{\rm max}$ rises above 2. 
 
 The red dashed lines in Fig.~\ref{fig:betaMAX} are for constant ratios, $\zeta_{-16}/I_{\rm UV}$ from 1 to 100 (assuming $C=1$ in Eq.~[\ref{eq:beta}]). These lines show that unless CR attenuation is neglected, $\zeta_{-16}/I_{\rm UV}$ must be very large for cosmic-rays to ever affect the positions of the HI-to-H$_2$ transition points.  In particular, for characteristic Galactic conditions, with $\zeta_{-16}/I_{\rm UV}\approx 1$ and ${\tilde \sigma}_g=1$, this never occurs.  For example, for $\alpha G=2$, Eq.~(\ref{eq:betaM}) gives $\beta_{\rm max}=7$, so that 
 for $I_{\rm UV}=1$, the free-space cosmic-ray ionization rate must be larger than $3\times 10^{-15}$~s$^{-1}$ for the transition point to be affected. Even without CR attenuation the ionization rate would have to be very large, and greater than $9\times 10^{-16}$~s$^{-1}$.




\label{lastpage}
\end{document}